\documentclass[final,3p,times,twocolumn,authoryear]{elsarticle}
\usepackage{amssymb}
\usepackage{amsmath}
\usepackage{lineno}
\usepackage[abs]{overpic}
\usepackage{upgreek} % for non-italic greek symbols
\usepackage{svg}
\usepackage{booktabs}
\usepackage{hyperref}
\usepackage{xcolor}       % For defining and using custom colors
\usepackage{siunitx}    % units package
\DeclareSIUnit\pixel{pixel}
\usepackage{acronym}    % allows consistent definition of acronyms used throughout the text
\acrodef{AER}[AER]{Address-Event Representation}
\acrodef{BL}[BL]{Boundary Layer}
\acrodef{CD}[CD]{Contrast Detection}
\acrodef{CFD}[CFD]{Computational Fluid Dynamics}
\acrodef{DMD}[DMD]{Dynamical Mode Decomposition}
\acrodef{DNS}[DNS]{Direct Numerical Simulation}
\acrodef{DR}[DR]{Dynamic Range}
\acrodef{DVS}[DVS]{Dynamic Vision Sensing}
\acrodef{DSR}[DSR]{Dynamic Spatial Range}
\acrodef{DVR}[DVR]{Dynamic Velocity Range}
\acrodef{EBIV}[EBIV]{Event-Based Imaging Velocimetry}
\acrodef{EBV}[EBV]{Event-Based Vision}
\acrodef{FOV}[FOV]{Field of View}
\acrodef{FPGA}[FPGAs]{Field-Programmable Gate Arrays}
\acrodef{FWHM}[FWHM]{Full-Width-Half-Mean}
\acrodef{HFSB}[HFSB]{Helium Filled Soap Bubbles}
\acrodef{HS}[HS]{High-Speed}
\acrodef{HWA}[HWA]{Hotwire Anemometry}
\acrodef{LDA}[LDA]{Laser Doppler Anemometry}
\acrodef{LES}[LES]{Large-Eddy Simulation}
\acrodef{LOR}[LOR]{Low-Order Reconstruction}
\acrodef{LPT}[LPT]{Lagrangian Particle Tracking}
\acrodef{PDF}[PDF]{Probability Density Function}
\acrodef{PIV}[PIV]{Particle Image Velocimetry}
\acrodef{POD}[POD]{Proper Orthogonal Decomposition}
\acrodef{ppp}[ppp]{particles per pixel}
\acrodef{PSD}[PSD]{Power Spectral Density}
\acrodef{PTV}[PTV]{Particle Tracking Velocimetry}
\acrodef{PWM}[PWM]{Pulse Width Modulation}
\acrodef{ROM}[ROM]{Reduced-Order Modeling}
\acrodef{RMS}[RMS]{Root Mean Square}
\acrodef{RMSE}[RMSE]{Root Mean Square Error}
\acrodef{2D-2C-PIV}[2D-2C-PIV]{two-dimensional (planar), two component particle image velocimetry}
\acrodef{ROI}[ROI]{Region of Interest}
\acrodef{SVD}[SVD]{Singular Value Decomposition}
\acrodef{TBL}[TBL]{Turbulent Boundary Layer}
\acrodef{TKE}[TKE]{Turbulent Kinetic Energy}

\usepackage{tikz}
\newcommand*\circled[1]{\tikz[baseline=(char.base)]{
    \node[shape=circle, draw=white, text=white, inner sep=2pt, very thick] (char) {#1};}}

\newcommand*\circledtext[1]{\tikz[baseline=(char.base)]{
            \node[shape=circle,draw,inner sep=2pt] (char) {#1};}}

\usepackage{graphicx} % To include graphics
\usepackage{subcaption} % For subfigures
\usepackage[normalem]{ulem}
\journal{Experimental Thermal and Fluid Science}

\begin{document}

\begin{frontmatter}

\title{An assessment of event-based imaging velocimetry for efficient estimation of low-dimensional coordinates in turbulent flows.} %% 

\author[label1]{Luca Franceschelli} %% Author name
\author[label2]{Christian E. Willert}
\author[label1]{Marco Raiola}
\author[label1]{Stefano Discetti}

\affiliation[label1]{organization={Department of Aerospace Engineering, Universidad Carlos III de Madrid},%Department and Organization
            addressline={Avda. Universidad 30}, 
            city={Leganés},
            postcode={28911}, 
            state={Madrid},
            country={Spain}}

\affiliation[label2]{organization={DLR Institute of Propulsion Technology, German Aerospace
Center},%Department and Organization
            addressline={Linder Höhe}, 
            city={Köln},
            postcode={51170}, 
            country={Germany}}

%% Abstract
\begin{abstract}
This study explores the potential of neuromorphic \ac{EBV} cameras for data-efficient representation of low-order model coordinates in turbulent flows. Unlike conventional imaging systems, \ac{EBV} cameras asynchronously capture changes in temporal contrast at each pixel, delivering high-frequency output with reduced data bandwidth and enhanced sensitivity, particularly in low-light conditions.
Pulsed \ac{EBIV} is assessed against traditional \ac{PIV} through two synchronized experiments: a submerged water jet and airflow around a square rib in a channel. The assessment includes a detailed comparison of flow statistics and spectral content, alongside an evaluation of reduced-order modeling capabilities using \ac{POD}.
The event stream from the \ac{EBV} camera is converted into pseudo-snapshots, from which velocity fields are computed using standard \ac{PIV} processing techniques. These fields are then compared after interpolation onto a common grid.
%Results indicate that \ac{EBIV} effectively reproduces flow statistics and spectral content, although discrepancies in the  \ac{PSD} are noted at high frequencies, attributed to higher noise levels in the \ac{EBIV} data. The air channel flow, in particular, presents more complexities, resulting in appreciable discrepancies in second-order statistics. Nevertheless, \ac{EBIV} captures the majority of the spectral content accurately.
Modal analysis demonstrates that \ac{EBIV} can successfully identify dominant flow structures, along with their energy and dynamics, accurately discerning singular values, spatial modes, and temporal modes. While noise contamination primarily affects higher modes—less critical for flow control applications—overall performance remains robust. Additionally, comparisons of \ac{LOR} validate EBIV’s capability to provide reliable reduced-order models of turbulent flows, essential for flow control purposes. These findings position EBV sensors as a promising technology for real-time, imaging-based closed-loop flow control systems.
\end{abstract}

% \todo{add graphical abstract}

% \begin{highlights}
% \item Event-Based Vision enables real-time velocimetry with less data.

% \item Statistics and spectral content matches well with conventional PIV.

% \item Modal analysis shows that EBIV accurately captures dominant flow structures.

% \item EBIV demonstrates potential for real-time estimation of low-order coordinates.
% \end{highlights}

%% Keywords
\begin{keyword}
PIV
\sep event-based velocimetry
\sep modal decomposition
\sep POD
\sep dimensionality reduction
\sep flow control
\end{keyword}

\end{frontmatter}

% \linenumbers

\section{Introduction}\label{Sec:Intro}
\noindent
\Acf{PIV} is recognized as a well-established tool in fluid mechanics, offering quantitative flow field measurements and contributing to a deep understanding of flow behavior. Its ability to capture instantaneous velocity fields enables researchers to analyze complex flow dynamics effectively, which has made it a fundamental diagnostic tool in both experimental fluid mechanics and engineering applications \citep{PIVBookRaffel:2018}. 
A key advantage of PIV lies in its ability to provide non-intrusive, full-field snapshot data of the time-varying flow, delivering critical insights into the flow's topological features, such as vortex structures, flow separations, and other patterns essential for understanding the flow dynamics.
This capability makes it particularly attractive for flow control applications, as it has the potential to enhance system observability, that is, how accurately the complete internal state of the flow can be inferred from sensor data. Enhanced observability allows for more accurate state estimation, enabling the design of more effective feedback control strategies and improving the ability to predict, manipulate, and stabilize complex flow behaviors \citep{brunton2015closed, duriez2017machine}.

However, the complexity and processing time associated with imaging techniques—both in image capture and analysis—have hindered their use as real-time "sensors" for fluid flow control. Previous studies have explored this potential, achieving time delays on the order of $100\,\text{ms}$ from image acquisition to processing \citep{willertRT, siegel2003}. More recently, optical flow algorithms \citep{quenot1998particle,liu2008fluid} have emerged as promising online imaging-based velocimetry techniques for flow control. Nevertheless, the computation frequency remains below $100\,\text{Hz}$ in water and even lower in air, despite efforts in slendering the processing at the expense of accuracy \citep{gautier2015closed, varon2019adaptive}. 

In specific flow scenarios characterized by well-defined flow configurations, computational efficiency can be markedly improved by narrowing the focus of analysis. Strategies such as focusing on a 'selected region of interest' (\ac{ROI}), using single-line \ac{PIV}, or adopting 'coarse-grained' \ac{PIV} leverage prior knowledge of the flow structures position and characteristics to reduce computational complexity while maintaining sufficient accuracy for control purposes \citep{braud2018real,kanda2022proof}. 
However, these methods are inherently tailored to particular applications and may not capture the full complexity of turbulent flows in more generalized scenarios. Furthermore, despite the reduction in computational time, these techniques still might fail to meet the requirements of actual industrial applications.

The primary source of these delays is identified as the image acquisition process, where the bottleneck lies in the large data size generated during imaging. For instance, consider a high-speed camera capturing images at a frequency of 1 kHz with a relatively small 1-megapixel sensor. The output reaches approximately \(10^3 \times 10^6 = 10^9\) samples per second, where each sample corresponds to the information from a single pixel. This vast amount of data must then be processed and validated through velocimetry algorithms, with control decisions based on the velocity output needing to be made within a reasonable time-frame. 

One method for enhancing real-time imaging velocimetry is the use of dedicated hardware, such as \ac{FPGA}. \ac{FPGA} allow for on-the-fly image processing, bypassing the computational delays typically associated with software-based algorithms. For example, hardware-accelerated PIV systems have demonstrated the capability to perform real-time image correlation and vector computation, achieving frame rates and data throughput well beyond what is feasible with general-purpose computing \citep{kreizer2010real,zhao20203d,Ouyang:RTPIV:2022}. These systems are particularly advantageous in applications requiring high-speed feedback control or large-scale flow analysis, as they offload computation-intensive tasks directly to optimized hardware.
However, \ac{FPGA}-based systems, while powerful, require specialized programming and are constrained by fixed hardware architectures, limiting their flexibility for diverse imaging setups or evolving algorithmic requirements.
Importantly, these approaches, including hardware-based accelerations, selective \ac{ROI} techniques, and algorithmic optimizations, are not mutually exclusive. They can be combined to achieve even higher processing speeds, offering complementary solutions to address various aspects of the real-time imaging bottleneck.\\
Thus, while emerging techniques and hardware solutions have shown promise in addressing the bottlenecks of image acquisition and processing, achieving real-time flow control across a wide range of applications remains an ongoing challenge requiring further research and development.

Lastly, conventional high-speed CMOS cameras typically do not permit continuous data streaming over extended periods and require the use of high-energy light sources, further complicating the setup.

Flow control applications often rely on high-frequency single-point sensing techniques such as thermal anemometry, laser Doppler velocimetry, and fluctuating pressure measurements with microphones. These methods are capable of measuring flow properties at frequencies extending into the kHz regime while maintaining a manageable data rate for long acquisition times \citep{dacome2024opposition, audiffred2024reactive}. However, both thermal anemometry and hydrodynamic pressure-based measurements are inherently intrusive, potentially perturbing the flow. Additionally, most of high-speed techniques available provide point measurements and thus requiring the global state of the flow to be inferred from localized single-point data.

% Neuromorphic sensors, also referred to as \acf{EBV} or \acf{DVS}, provide a technological novelty, with the potential of significantly reducing the inherent time delay introduced by the image acquisition and processing of conventional cameras.
In this context neuromorphic sensors, also referred to as \acf{EBV} or \acf{DVS}, have been gaining significant attention across various engineering applications due to the paradigm shift they bring to the image acquisition process and the range of advantages they offer. In real-time imaging scenarios, these sensors hold the potential of significantly reducing the inherent time delays associated with image acquisition and processing in conventional cameras.
Whereas in conventional framing cameras exposure and read-out times are globally controlled and set for the whole sensor, each pixel of an \ac{EBV} sensor operates as an independent asynchronous temporal contrast change detector. A pixel triggers an event only when the integral of the temporal contrast, represented as the temporal variation of the logarithmic intensity 

\begin{equation}
    C_\text{temporal} = \frac{1}{I(t)}\frac{dI(t)}{dt} = \frac{d}{dt} \ln[I(t)],
\end{equation}

\noindent surpasses a predefined threshold in comparison to the previous event \citep{Lichtsteiner:2008,Posch:2014}. In this context, $I(t)$ denotes the photo-current generated by the photodiode, which is proportional to the pixel illumination. 
The sensor output is a continuous asynchronous stream of events, consisting of the activated pixel address, the time-stamp of the event referred to the global clock of the sensor and the polarity, indicating whether the event is a positive (ON) or negative (OFF) contrast change. 
The event-triggering mechanism and its relation to logarithmic intensity changes enable these sensors to achieve a sensor \ac{DR} of approximately 120 dB. The \ac{DR}, which quantifies the ability to capture details in both bright and dark regions of the field of view simultaneously, is defined in decibels as the ratio of the maximum measurable signal to the noise floor:
\begin{equation}
    \ac{DR}_{\text{dB}} = 20 \cdot \log_{10}\left(\frac{\text{Maximum signal}}{\text{Noise floor}}\right)
\end{equation}
For digital cameras commonly used for \ac{PIV}, the \ac{DR} is traditionally expressed in terms of bit depth (\ac{DR}$_{\text{N}}$) rather than decibels. The relationship between the two \citep{tomarakos2002relationship} can be approximated as:
\begin{equation}
    \ac{DR}_{\text{N}} = \frac{\ac{DR}_{\text{dB}}}{6}
\end{equation}
Using this conversion, \ac{EBV} sensors achieve an equivalent bit depth of 20 bits. In comparison, conventional high-speed cameras typically achieve a \ac{DR} of 60–70 dB, corresponding to 10–12 bits, while advanced sCMOS cameras with adequate cooling and low read noise can achieve higher dynamic ranges, up to 16 bits (96 dB), albeit at slower readout speeds.\\
This extended dynamic range makes event-based cameras particularly advantageous for imaging scenarios with high contrast, such as regions with simultaneous bright illumination and deep shadows—common in experimental fluid mechanics. Additionally, the high sensitivity of \ac{EBV} sensors at low-intensity levels broadens their applicability to situations where the available light is insufficient for conventional \ac{PIV} measurements.\\
Considering the temporal resolution of today's \ac{EBV} technology, in the order of \qty{100}{\us}, flow field events can be recorded at an analogous of a framing frequency of conventional high-speed CMOS-based framing cameras exceeding several kHz, but with a significantly reduced data bandwidth. The possibility of accessing volumetric flow field information at a sufficiently high frequency, with a ``slim''-enough stream of events to be easily and rapidly processed, together with the high sensitivity to contrast changes, paves the way to the possibility for PIV-based real-time closed-loop control in turbulent flows.

However, the use of these sensors is subject to certain limitations. The arbiter, a designated circuit element responsible for associating each event with the global sensor clock, can effectively process events only up to a maximum number of simultaneous events. This limitation consequently constrains the speed of the flow field case and the dimension of the resolved volume \citep{willert2022event}.
Furthermore, the event sensor's pixels are characterized by latency (i.e. response time), which hinders the temporal accuracy of event detection, resulting in information loss. In the context of \ac{PIV} applications, a notable loss of accuracy is observed due to the inherent impossibility of achieving sub-pixel accuracy, stemming from the binary event definition as ON/OFF states. To a certain extent, this can be mitigated using the equivalent of multiple frame \ac{PIV} processing schemes \citep{willertpuls}.
 
\cite{Gallego2022} present comprehensive insights into the current state of event-based vision, encompassing its diverse applications and the associated data processing landscape. In the field of fluid mechanics, \cite{willertpuls} successfully addressed challenges related to event rate limitations and sensor inherent time-latency by employing a pulsating laser as a light source. This development, known as pulsed event-based imaging velocimetry (\textit{pulsed-EBIV}), is implemented and referred to as EBIV in this work. The results, employing typical PIV-processing algorithms across various test cases, align well with measurements obtained using conventional \ac{PIV} cameras.
Given the asynchronous and independent operating principle of EBV sensors, \acf{PTV} approaches have also been explored. Recently, \cite{TrackAER} integrated a real-time 3D event-based \ac{PTV} acquisition system into their wind tunnel, demonstrating its capability to reconstruct particle trajectories across a range of industrial applications. Additionally, stereoscopic PTV methods \citep{Wang:StereoEBPTV:2020} and optical-flow-based approaches leveraging EBV sensors have been investigated \citep{BOS}.
More recently, \cite{willert2024dynamic} implemented a 3D \ac{LPT} algorithm on event-based vision (EBV) data to characterize the near-wall behavior of a turbulent boundary layer in air. This approach achieved data quality comparable to that of state-of-the-art high-speed framing cameras.

To advance towards imaging-based flow control, it is essential to address not only the imaging technique but also dimensionality reduction. The amount of information obtained from imaging velocimetry is typically very large, while often flow control problems aim at identifying only a handful relevant coherent structures or patterns in the flow field \citep{rowley2017model}. Such features can be effectively represented through \ac{ROM} techniques that utilize only a few latent coordinates. Dimensionality reduction methods, such as \acf{POD} \citep{berkooz1993proper} and Dynamic Mode Decomposition \citep{schmid2022dynamic}, can provide a description of the main flow features, significantly reducing the problem's dimensionality and minimizing computational demands for flow control \citep{brunton2015closed, taira2017modal}.

Furthermore, EBV-based velocity measurements are generally more noise-contaminated compared to conventional \ac{PIV}/\ac{PTV} results, which can affect the latent coordinates in \ac{ROM}. However, this degradation, being distributed over the full rank of latent coordinates, primarily impacts the higher-order ones. As a result, \ac{ROM} techniques can serve as effective de-noising tools for reconstructing velocity fields \citep{epps2010error,raiola2015piv, brindise2017proper,epps2019singular,gu2024denoising}. 

In view of the future application of EBV cameras as sensors for closed-loop flow control, this study aims to assess the reduced-order modeling capabilities of EBV cameras, comparing them with conventional frame cameras. The key idea is that, since flow control strategies can be targeted primarily to low-order latent variables (e.g., the first modes in \ac{POD}), the potential reduction in quality observed with EBIV measurements compared to PIV should have minimal impact on potential control performance.

Considering an example of future application leveraging \ac{ROM} for flow control, a spatial modes basis, computed via \ac{POD} can be first defined offline using a preliminary set of velocity snapshots. This basis represents the dominant flow structures and allows for the projection of instantaneous velocity fields to extract the corresponding temporal mode coefficients, i.e. the coordinates in the latent space. Provided the projection process is fast, this methodology enables real-time identification of flow dynamics while leveraging the reduced computational burden afforded by the latent coordinate representation. In the present analysis, the same approach has been applied, with the difference that the velocity snapshots and projection computations were all performed off-line. Nonetheless, the feasibility of this framework for real-time applications is demonstrated by its computational efficiency and alignment with operational constraints in imaging-based flow control.

Results from EBIV and conventional PIV camera, hereafter referred to simply as \textit{PIV}, are compared using two experimental datasets, where data are acquired synchronously for both cameras: a submerged water jet flow and a channel airflow with a squared rib obstacle. The resulting flow statistics, spectral content, and modes from \ac{POD} (snapshot method, \citealt{sirovich1987turbulence}) are compared.

%%%%%%%%%%%%%%%%%%%%%%%%%%%%%%%%%%%%%%%%%%%%%%%%%%%%%
\section{Experimental Datasets}\label{sec:expset}
The comparative analysis between \ac{PIV} and \ac{EBIV} was conducted using two different experimental datasets, produced for the purpose: a submerged water jet flow and a channel air flow with a spanwise square rib. In both cases, synchronized acquisition was performed using a CMOS sensor camera and an EBIV camera, positioned facing each other.
The \ac{FOV} and image resolution $Res$ - defined in terms of pixel/mm, were closely matched to ensure an objective comparison of the results obtained from both systems, as well as consistent PIV processing. 
Also, the co-axial viewing arrangement ensured matching scattering behaviour of the observed particle fields for both cameras.
In both measurement configurations, EBIV measurements were performed using the EBV camera Prophesee EVK4, featuring the Sony IMX636 sensor with a resolution of $1280 \times 720$ pixels and a square pixel pitch of $4.86\,\upmu\text{m}$. 
Table \ref{tab:comparison} summarizes the main features of the two experimental setups.
% Under nominal conditions the pixel latency is below $100\,\upmu\text{s}$.

\begin{table}[]
\centering
\caption{Comparison of experimental parameters for the jet flow and square rib cases. *Resolution here refers to a rounding value between the two used cameras. Specific values are provided in section \ref{sec:expset}.}
\renewcommand{\arraystretch}{1.5} % Increase the interline spacing
\begin{tabular}{lcc}
\toprule
& \textbf{Jet Flow} & \textbf{Square Rib} \\ 
\midrule
Medium       & Water & Air  \\ 
$U_\infty$ [m/s]       & 0.13 & 2.8  \\ 
Characteristic \\ length [mm]                & $D = 20$ & $H = 8.18$ \\ 
% $Re$                    & - & 14000 \\ 
$Re_l$                  & 2600 & 1500 \\ 
$f$ [Hz]                & 100 & 5000 \\ 
Pulse width [$\upmu$s]    & 1000 & $<$0.2 \\ 
$Res$* [pixel/mm]          & 9.2 & 30.3 \\ 
Record duration [s]     & 40 & 6 \\ 
$N_t$ [snapshots]           & $3.5\cdot 10^3$ & $3\cdot 10^4$ \\ 
% $N_p$ [gridpoints]           & 
% $12312$ 
% & \begin{tabular}[c]{@{}c@{}}
% $2976$ \\
% $4836$ \\
% $4648$
% \end{tabular} \\
IW size [pixel]   & $32\times 32$ & $48\times 48$ \\
PIV data rate [MB/s]   & 210 & 11500 \\
EBIV data rate [MB/s]   & 130 & 75 \\
Events(p=1)/snapshot  & 210000 & 3000 \\
\bottomrule
\end{tabular}
\label{tab:comparison}
\end{table}

\subsection{Submerged water jet flow}
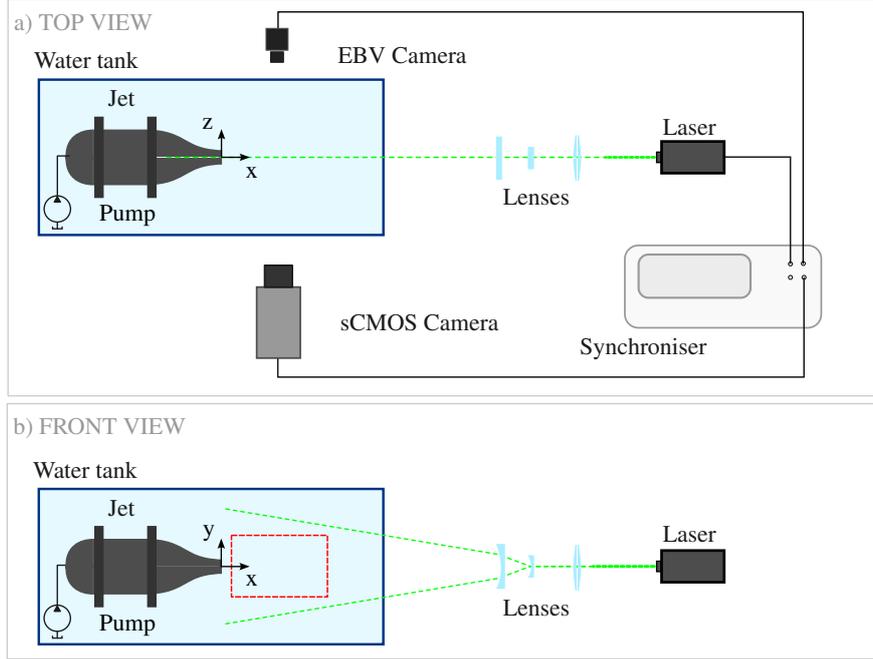
\begin{figure*}[]
    \centering
    \begin{overpic}[width=0.7\linewidth]{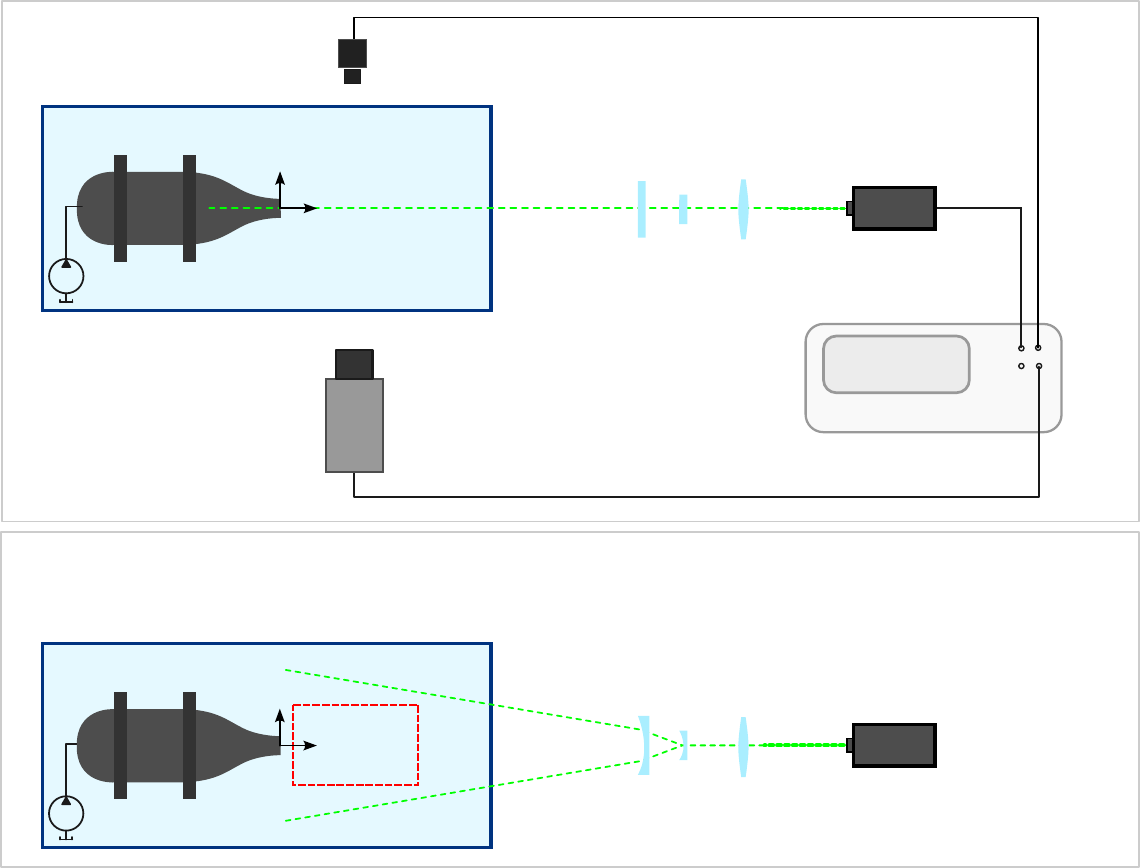}
        % Example annotations:
        \put(5,237){(a) Top View}
        \put(12,222){Water Tank}
        \put(115,230){EBV Camera}
        \put(39,208){Jet}
        \put(30,165){Pump}
        \put(75,205){z}
        \put(90,180){x}
        \put(120,125){sCMOS Camera}
        \put(185,170){Lenses}
        \put(245,200){Laser}
        \put(230,114){Synchronizer}
        
        \put(5,85){(b) Side View}
        \put(12,68){Water Tank}
        \put(39,52){Jet}
        \put(30,10){Pump}
        \put(75,50){y}
        \put(90,25){x}
        \put(186,15){Lenses}
        \put(245,46){Laser}
    \end{overpic}
    \caption{Submerged water jet-flow, sketch of the experimental set-up: in red the common region onto which the obtained velocity fields are interpolated.}
    \label{fig:exp}
\end{figure*}
The jet nozzle, with an exit diameter $D=20$ mm, is operated at a bulk velocity of approximately $U_b \approx 0.13\, \text{m/s}$, resulting in a turbulent jet flow at a Reynolds number of $Re \approx 2600$. %\\
The jet is contained within an $80 \times 60 \times 40\, \text{cm}^3$ water tank and seeded with polyamide particles of $56\,\upmu \text{m}$ diameter. Illumination on the symmetry plane of the jet is provided by a low-cost, pulse-width modulated laser, originally designed for wood engraving (LaserTree LT-40W-AA), with a power of $5\,\text{W}$. The laser beam is shaped into a thin laser sheet through a set of lenses, reaching a thickness of around $1-2\,\text{mm}$ at the imaged \ac{FOV}. The laser is controlled with a pulse generator:  the pulse width is set at $1\,\text{ms}$, and the time interval between two consecutive laser pulses is $dt = 10\,\text{ms}$, corresponding to a frequency $f = 100\,\text{Hz}$. The pulse generator also provides the acquisition trigger signal for the PIV camera and the reference trigger signal for the event-based camera. In the event-based camera, this signal serves as a reference point within its data stream, ensuring temporal alignment between the two camera systems and thereby facilitating coherent data acquisition and analysis.
The experimental setup is illustrated in Fig.~\ref{fig:exp}.

For the conventional PIV image acquisition, the flow field is captured using an sCMOS camera (Andor Zyla) with a 5.5 megapixels sensor and a pixel pitch of \qty{6.5}{\um}.
The resolution values for the conventional PIV camera and the EBV camera are $Res_\text{PIV} = 9.25\,\text{pixel/mm}$ and $Res_\text{EBIV} = 9.13\,\text{pixel/mm}$, respectively. The sCMOS camera field of view has been cropped to a resolution of $1500 \times 720$ pixels, covering a domain of approximately $8.1 \times 3.9 D$.
The EBIV imaged flow field covers a domain of approximately $140\times 80\, \unit{\square\mm}$, corresponding to $7 D \times 4 D$. For the sake of comparison, the obtained velocity fields are interpolated onto a common grid, spanning $0.5\leq x/D\leq 6.5$ and $-1.75\leq y/D\leq 1.75$, shown as the red dashed box in Fig.~\ref{fig:exp}a. The acquisition software allows for user-defined tuning of event sensitivity. However, for the purpose of providing general considerations about this new technology, no tuning has been implemented, leaving room for potential improvements in the quality of the acquired data.  

The acquisition consists of an event stream spanning several minutes. The acquired data stream has an average event rate of approximately $60\times 10^6\,\text{Ev/s}$, corresponding to a data rate of roughly $130\,\text{MB/s}$, where the positive events corresponds to only the 35\,\% of the total. In comparison, the data rate for sCMOS raw image acquisition, without any compression, is approximately $210\,\text{MB/s}$. While this represents only about a factor of 2 difference, the sCMOS data rate can be further reduced—potentially by an order of magnitude—using image compression or binary thresholding techniques, making the gap between the two approaches less significant.
A time-resolved sequence consisting of $N_t = 3.5\times 10^{3}$ samples is captured.

\subsection{Channel flow with square rib} 
\label{subsec:channel}

\begin{figure*}[]
    \centering
     \begin{overpic}[width=1.5\columnwidth,unit=1mm]{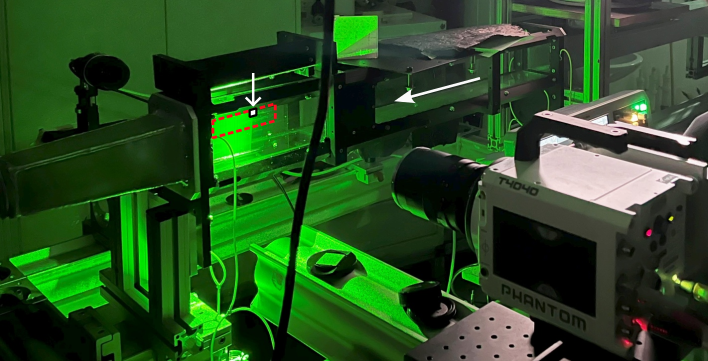}
          \put(56,40){\circled{1}}
          \put(4,47.5){\circled{2}}
          \put(87,15){\circled{3}}
          \put(39,50){\circled{4}}
          \put(28,37){\circled{5}}
          % \put(36,32){\circled{6}}
          \put(70,47){\color{white} \( U_\infty \)}
     \end{overpic}

    \caption{Photograph of the channel flow setup with HS-PIV camera in foreground and \ac{EBV} camera on opposite side.\\
    \protect\circledtext{1} Channel flow test section; \protect\circledtext{2} EBV camera;\protect\circledtext{3} HS-PIV camera; \protect\circledtext{4} square rib; \protect\circledtext{5} overall \ac{FOV} from the multiple acquisitions.
    }
    \label{fig:channelflow_exp}
\end{figure*}
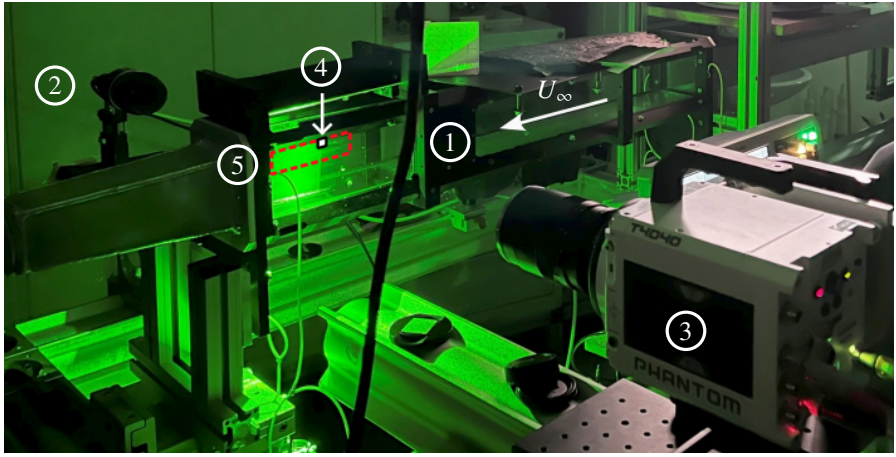

The channel flow experiment was conducted in the laboratory of the Institute of Propulsion Technology at the German Aerospace Center (DLR) in Cologne. 
A configuration similar to that used in \cite{willertpuls} was reproduced and depicted in Fig.~\ref{fig:channelflow_exp}. 
A spanwise square rib with side dimension $H = 8.18\,\text{mm}$ was placed on the upper surface of the channel flow facility which featured a square test section of $W = 76\,\text{mm}$. 
The air flow was seeded with \qty{1}{\um} paraffin droplets which were injected in the upstream settling chamber of the channel flow facility.
Light sheet illumination was realized using a  high-speed pulsed laser (Innolas/Iradion Nanio Air 532-10-V-SP) which provided short laser pulses of $<\qty{50}{\ns}$ duration at a rate of \qty{5}{\kHz}.

Several flow velocities corresponding to three primary flow regimes—laminar, transitional, and turbulent—around the rib were tested.
In this paper, only the results for the turbulent case are presented, as it is the most challenging yet still provides comparable results to the other flow conditions. In this case, the free-stream velocity is $U_{\infty} \simeq 2.8\,\text{m/s}$, yielding a bulk Reynolds number of $Re_W \simeq 14\,000$ and a Reynolds number based on the rib dimension of $Re_H \simeq 1\,500$. 
\begin{figure*}[]
    \centering
    \begin{overpic}[width=1\linewidth]{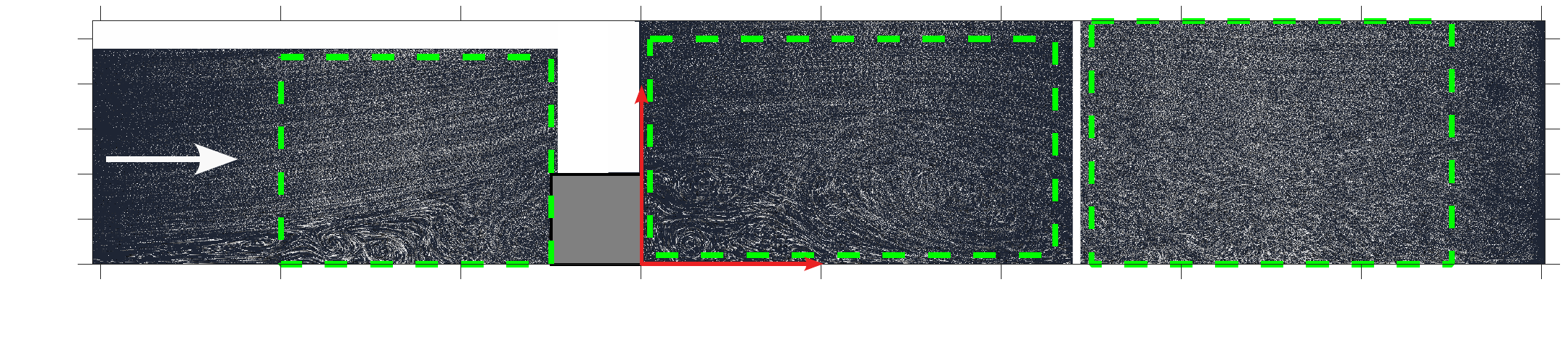}
        % Example annotations:
        \put(220,0){X/H}
        \put(0,50){\rotatebox{90}{Y/H}}
        \put(180,75){\color{red} Y}
        \put(240,32){\color{red} X}
        \put(40,60){\color{white} \( U_\infty\)}

        \put(14,21){0}
        \put(10,35){0.5}
        \put(14,48){1}
        \put(10,62){1.5}
        \put(14,75){2}
        \put(10,89){2.5}

        \put(24,10){-6}
        \put(79,10){-4}
        \put(133,10){-2}
        \put(190,10){0}
        \put(243,10){2}
        \put(296,10){4}
        \put(350,10){6}
        \put(404,10){8}
        \put(455,10){10}

    \end{overpic}
    \caption{Sketch of the resolved regions in the square rib experiment: the front, back, and far-back areas are outlined by green dashed lines. The X and y axes, along with the origin, are defined in red. All dimensions are normalized by the rib size, $H$. Background images are obtained from non-synchronized EBV camera acquisition, with an accumulation time of \qty{20}{\ms}}
    \label{fig:ribsketch}
\end{figure*}
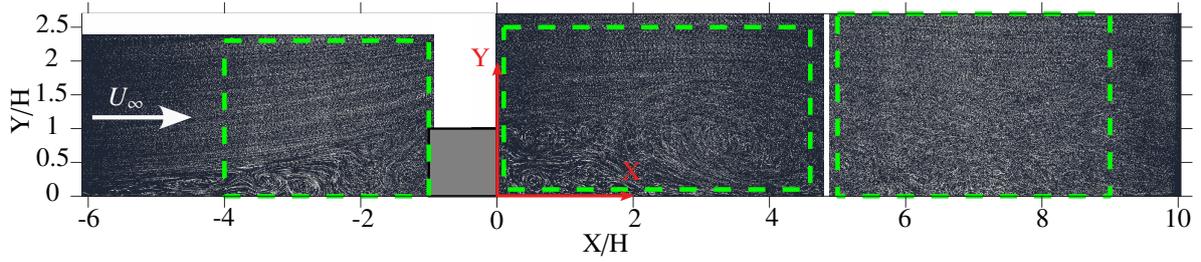

The \ac{HS} CMOS camera (HS-PIV camera in Fig.~\ref{fig:channelflow_exp}) used for this experiment is a Phantom T4040, featuring a 4.2 Mpixel sensor with a square pixel size of $9.27\,\upmu\text{m}$. For the measurements, the \ac{FOV} was cropped to a resolution of $1535\times 768$ pixels ($50 \times 25\,\unit{\square\mm}$) to approximately match the \ac{FOV} height of the \ac{EBV} camera. 
Operated at its full sensor resolution ($1280\times 720$ pixel), the \ac{FOV} of the \ac{EBIV} covered an area of $42 \times 24\,\unit{\square\mm}$, corresponding to approximately $5 \times 3\,\text{H}$.
To optimize the acquired event stream for PIV processing, the internal biases of the EBV camera were adjusted to decrease sensitivity to negative event triggers. This modification significantly reduced the occurrence of negative events, ensuring that most of the data bandwidth could be dedicated to positive events, thus preventing event overload. Additionally, the rib was consistently cropped out of the field of view. As highlighted in \cite{willertpuls}, the pulsating laser striking the anodized surface of the aluminum rib generates a considerable number of events, which reduces the sensor's capacity to register positive events before the arbiter saturates. 

The obtained data rate for the Phantom camera is around 11.5 GB/s. This is orders of magnitude higher than the one for EBIV (approximately 75 MB/s). EBIV recorded an average of $25\times 10^6$ Ev/s among which 55\% corresponded to positive ones.

Three distinct regions of the channel flow were investigated: the front, back, and far-back areas of the square rib. In dimensionless coordinates, the resolved areas correspond to $-4 \leq X/H \leq -1$, $0.1 \leq X/H \leq 4.6$, and $5 \leq X/H \leq 9$ in the longitudinal direction, respectively. In the direction orthogonal to the wall, the resolved areas are $0 \leq Y/H \leq 2.3$, $0.1 \leq Y/H \leq 2.5$, and $0 \leq Y/H \leq 2.7$, respectively. The origin is defined at the intersection of the wall and the back-facing side of the rib. As for the jet case, the obtained velocity fields were interpolated into the aforementioned common grids. Fig.~\ref{fig:ribsketch} shows non-synchronized EBIV images with an accumulation time of \qty{20}{\ms}, illustrating the origin definition and the resolved areas of the flow field, indicated by the green dashed lines. 

The resolution and relative alignment between the two cameras were computed using a common glass target with a printed chessboard pattern, positioned in the laser plane. Given the known distance between the squares on the chessboard, the resolution for each camera was automatically determined. On average, the PIV camera had a resolution of \(Res_{\text{PIV}} = 30.26\,\text{pixel/mm}\), while the EBV camera had a resolution of \(Res_{\text{EBIV}} = 30.31\,\text{pixel/mm}\). The use of the double-sided target also allowed for the evaluation of the angle between the two cameras. The resulting velocity fields could then be corrected for this angle, ensuring proper alignment. 

% The flow was seeded with **[Chris, please provide the type of particles]** particles, which had a diameter of approximately $1\,\upmu\text{m}$. The longitudinal mid-plane of the channel flow was illuminated with a pulsed laser operating at a frequency of $f = 5\,\text{kHz}$. The laser was directed from the bottom to the top and shaped into a thin plane using a set of lenses. \\
Similar to the jet flow case, the EBV camera received a trigger signal for synchronization with the Phantom camera. In this case, the signal was generated by the CMOS camera itself. 

Considering a convective time scale of $t_c = H/U_\infty \simeq 3\, \text{ms}$, a dataset covering at least 2000 $t_c$ was collected for statistical purposes, corresponding to $N_t = 30\,000$ velocity snapshots.
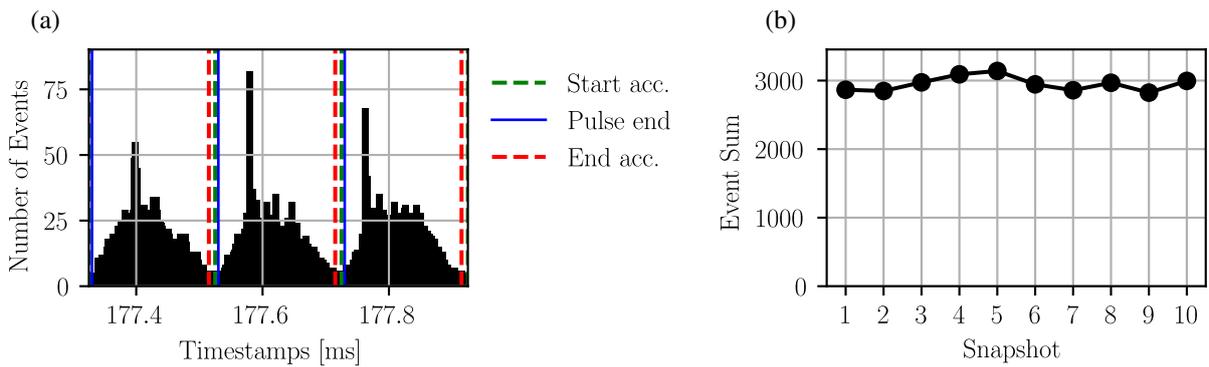
\begin{figure*}[]
    \centering
    \includegraphics{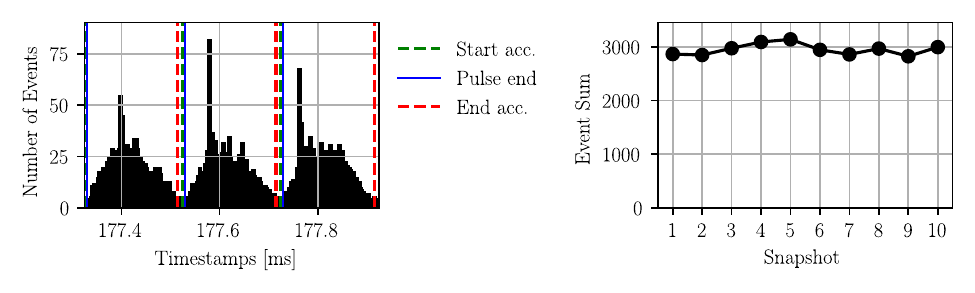}
    \begin{picture}(0,0)
        \put(-215,150){(a)}
        \put(60,150){(b)}
    \end{picture}
    \caption{Back of the square rib case: Example of positive event-rate from the EBV camera with a pulsating laser. (a) A histogram shows event bursts corresponding to the first three laser pulses. The green, blue, and red dashed lines indicate the beginning and end of the laser pulse, as well as the end of the accumulation time, respectively. (b) The total number of events accumulated for the first ten pseudo-snapshots is displayed. The first green line marks the initial trigger signal from the high-speed CMOS camera.}
    \label{fig:eventsrate}
\end{figure*}

\subsection{EBV camera data processing}
Following the procedure proposed by \cite{willertpuls} the event stream is partitioned to the reference trigger signal to obtain the set of $N_t$ pseudo-snapshots.
The obtained synthetic snapshots compress, on average, over $2 \times 10^5$ samples for the jet flow case, and around $3 \times 10^3$ samples for the square rib case— specifically, $3.1 \times 10^3$ for the front region case, $3 \times 10^3$ for the back region, and $5 \times 10^3$ for the far-back region.

Fig.~\ref{fig:eventsrate}a shows a histogram of the registered events over time for the square rib experiment. Green and blue dashed lines indicate the start and end of the laser pulse, respectively. A burst of positive contrast-change events (Polarity = 1) occurs immediately after the laser pulse illuminates the particles. A pseudo-image is then generated by accumulating all positive events registered from the beginning of the laser pulse over a user-defined \textit{accumulation time}, typically, the laser pulsing period. The high-frequency channel flow case shown in Fig.~\ref{fig:eventsrate} illustrates that the number of events does not drop to zero before the next laser pulse is fired. The accumulation time is therefore selected to capture the majority of the triggered events. In this case, the number of events not accumulated represents approximately 1\% of the total number of events triggered between two laser pulses, which are likely due to consistent noisy pixel activations, even when no light source is present. In contrast, in the case of jet flow, where the time interval between consecutive laser pulses is significantly larger, the behavior is much clearer. The event count decreases sharply and stays near zero during the period between the end of one laser pulse and the start of the next.

In some instances, very bright particles may trigger multiple activations of the same pixel within a single laser pulse. These multiple triggers are treated identically to single activations.

The resulting pseudo-images are binary, with pixel values of 0 indicating no activation and 1 indicating that the pixel was activated at least once.
\begin{figure*}[]
    \centering
    \begin{overpic}{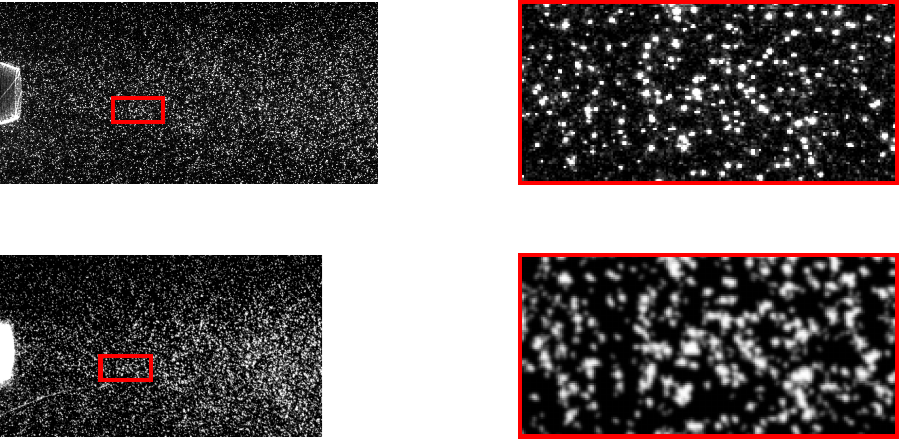}
        % Example annotations:
        \put(0,217){(a)}
        \put(250,217){(b)}
        \put(0,95){(c)}
        \put(250,95){(d)}
    \end{overpic}
    \caption{Comparison of input images for PIV processing: conventional PIV camera image (a,b) and EBV camera pseudo-snapshot (c,d). On the left, the full field of view (a,c). On the right, the zoomed red squared area (b,d). Both images correspond to the same laser pulse and have undergone pre-processing.}
    \label{fig:snaps}
\end{figure*}
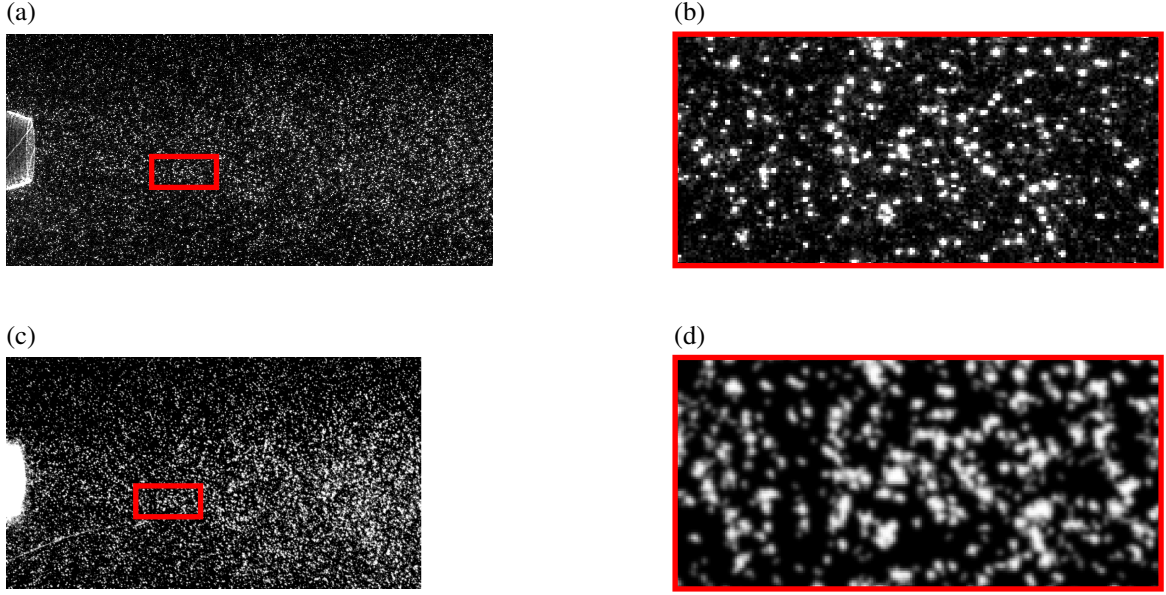

For the jet flow, characterized by larger particles and longer laser pulse duration, the raw pseudo-images display particle-like bodies with diameters spanning a few pixels. Single-pixel activations due to noisy events, which account for an average of $\sim 8\,\%$ of the total positive events, are removed using a convolutional filter. This filtering process is conceptually similar to morphological erosion in binary image processing, which is commonly used for noise reduction \citep{jamil2008noise}. In contrast, for the more challenging channel flow case, only single-activated pixels are generally observed, as the dimensions of the imaged particles on the sensor are typically on the order of 1–2 pixels, owing to the small size of the seeding particles. As a result, no clear distinction can be made between actual particles and noisy activations.

Subsequently, Gaussian smoothing with a kernel size of 0.75 pixel is applied to the raw images to prepare them for processing by standard PIV algorithms.
In Fig.~\ref{fig:snaps}, an example of snapshot from the jet experiment used in PIV algorithm processing is shown for both acquisition technologies. The pseudo-snapshot from the EBV camera (bottom) has been filtered as previously described. For the conventional PIV camera, cumulative minimum background removal was applied to the snapshot.

%astarita2005analysis, astarita2006analysis,
\subsection{PIV processing}
The raw image sequences were processed using PaIRS \citep{Paolillo2024}, an open-access PIV code developed at Università Federico II di Napoli. Matching PIV processing parameters were applied to both CMOS and EBV camera images for each experiment.
For the jet flow case, an iterative multi-grid/multi-pass image deformation algorithm was used, with final interrogation windows of $32 \times 32\,\text{pixel}$ and a 75\,\% overlap. In the channel flow case, the final interrogation windows were $48 \times 48\,\text{pixel}$, also with a 75\,\% overlap.

The use of simple 2-frame cross-correlation processes was chosen to evaluate real-time applications. For EBIV, significantly better results can be achieved when applying a multi-frame algorithm, particularly in high-speed velocimetry cases like the square rib experiment.
For the same reason, no post-processing filtering has been applied to the velocity fields, if not standard validation criteria during the PIV processing.

\begin{figure*}[]
    \centering
    % Top left figure
    \begin{subfigure}[t]{0.45\textwidth}
        \centering
        \includegraphics{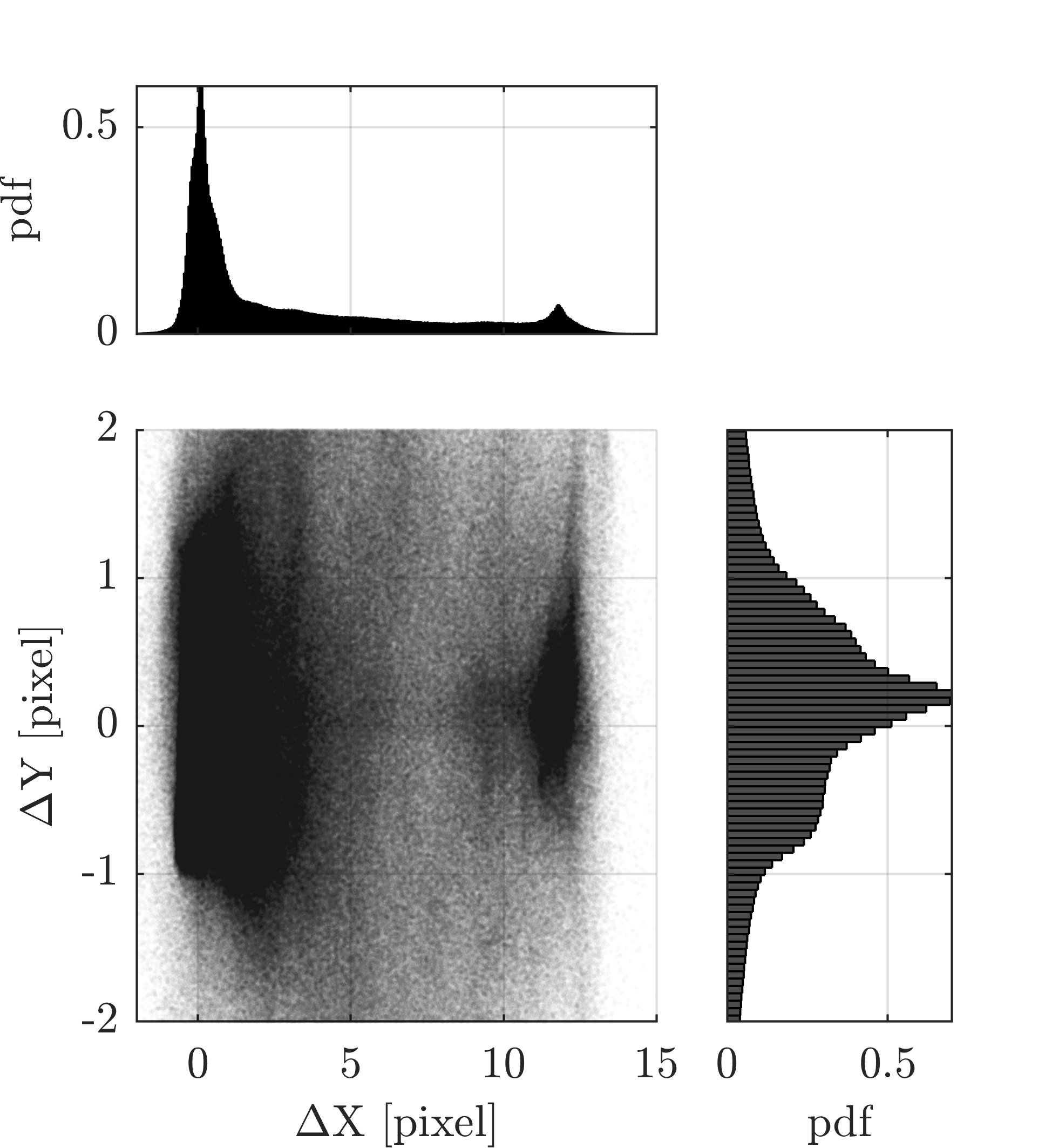} % replace with your figure
        % \caption{Water jet flow - PIV}
        \label{fig:subfig1}
    \end{subfigure}
    % Top right figure
    \begin{subfigure}[t]{0.45\textwidth}
        \centering
        \includegraphics{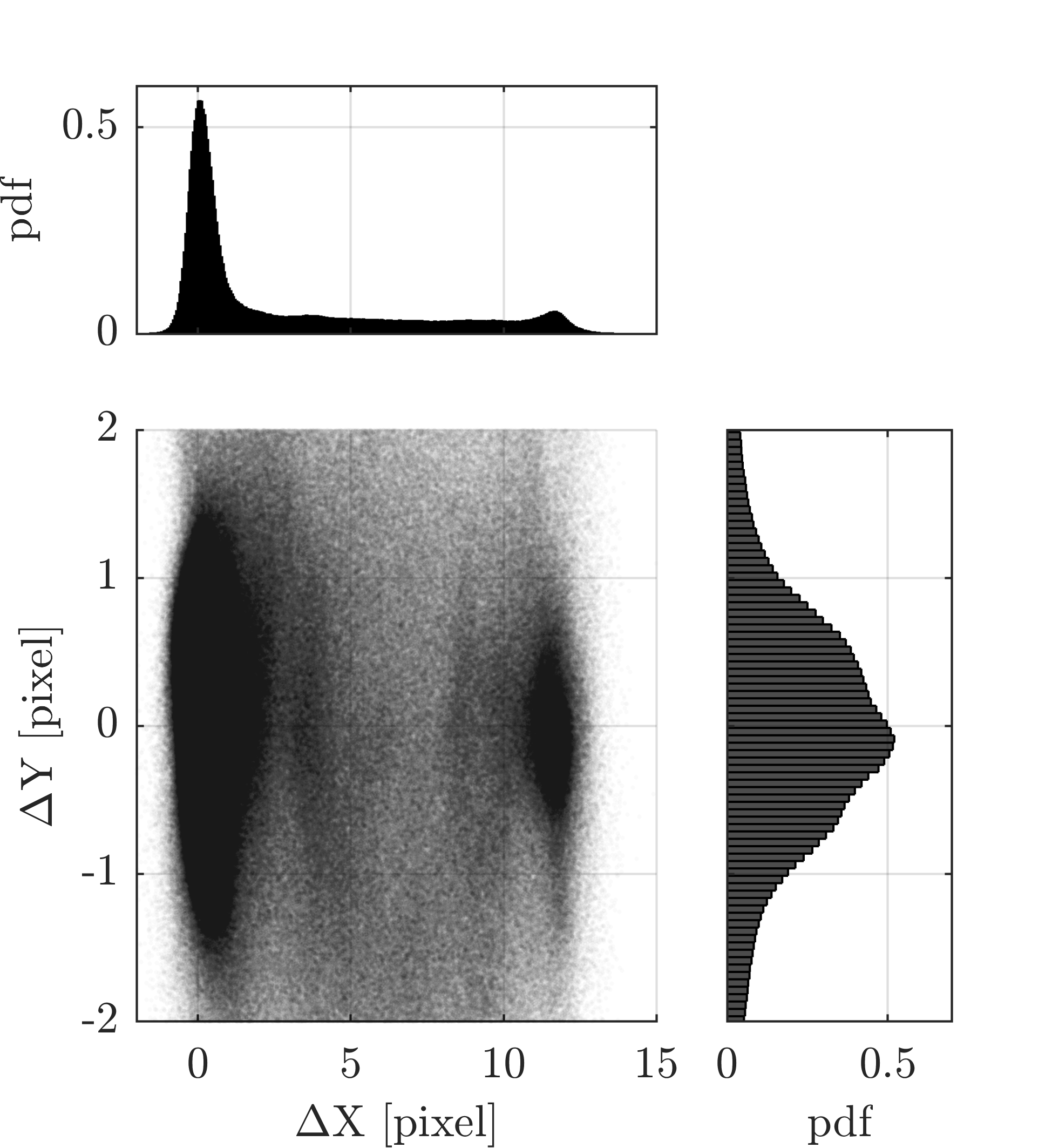} % replace with your figure
        % \caption{Water jet flow - EBIV}
        \label{fig:subfig2}
    \end{subfigure}
    \vskip\baselineskip % Create vertical space between rows
    % Bottom left figure
    \begin{subfigure}[t]{0.45\textwidth}
        \centering
        \includegraphics{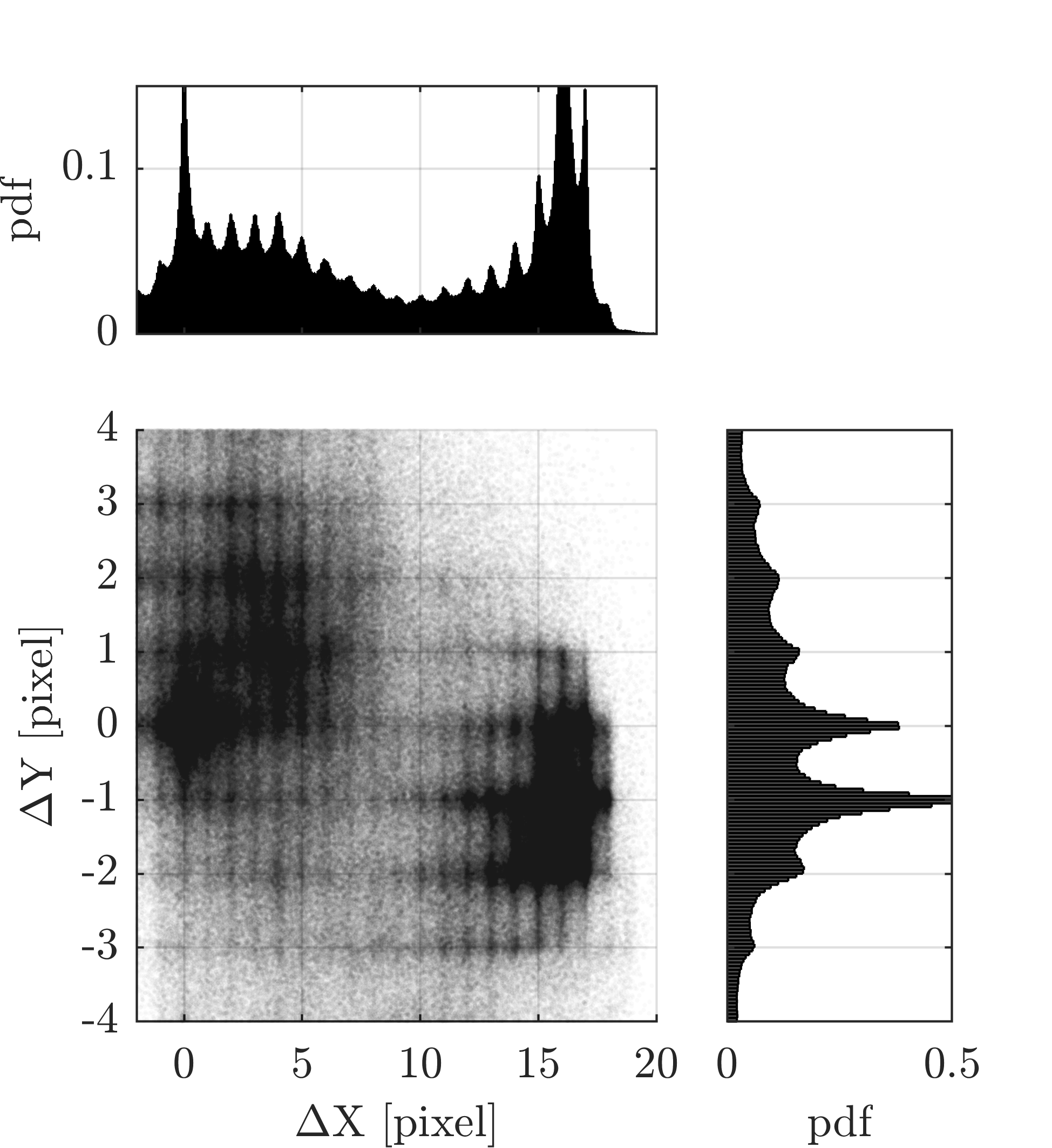} % replace with your figure
        % \caption{Channel flow with square rib (far-back region) - PIV}
        \label{fig:subfig3}
    \end{subfigure}
    % Bottom right figure
    \begin{subfigure}[t]{0.45\textwidth}
        \centering
        \includegraphics{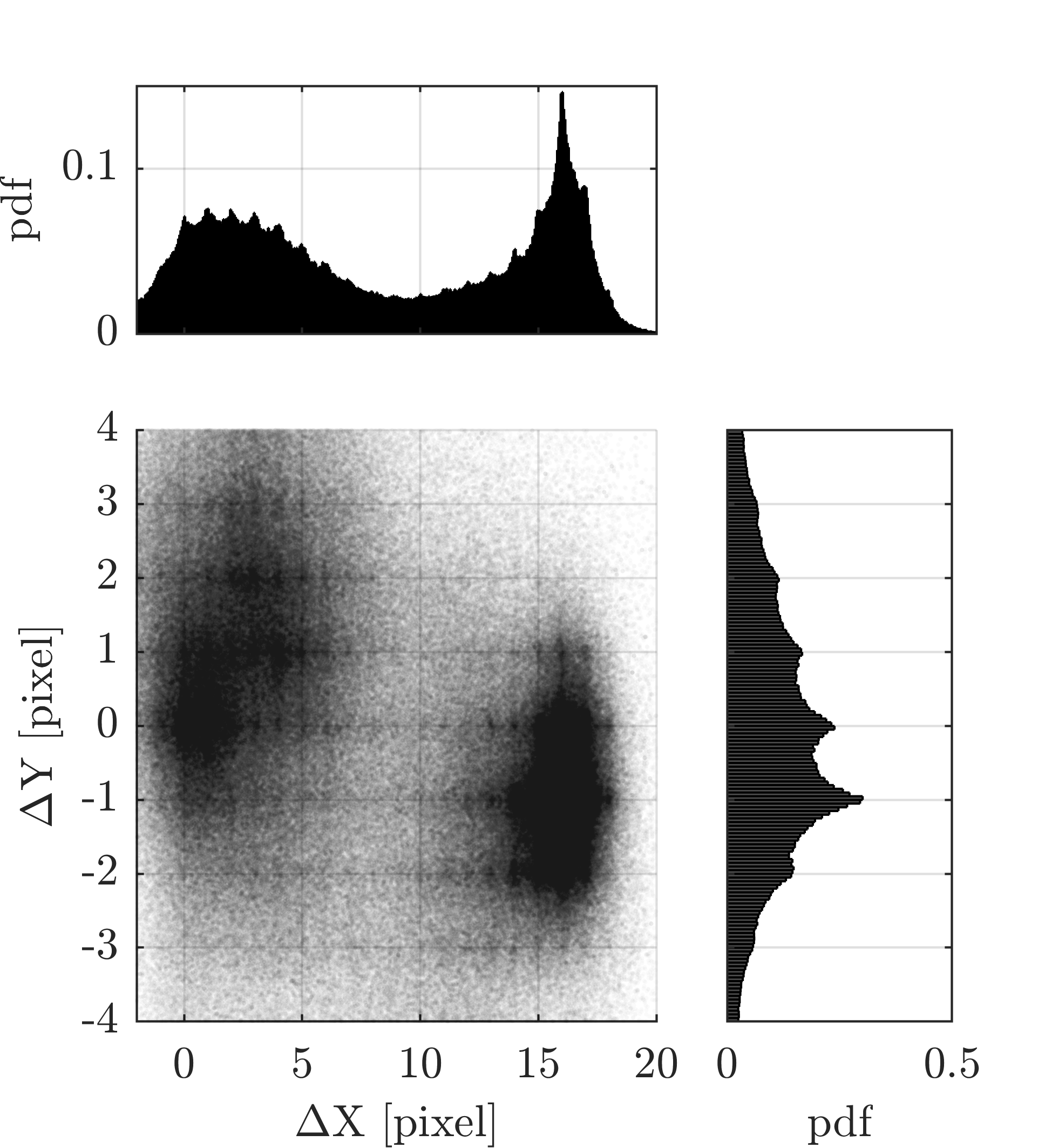} % replace with your figure
        % \caption{Channel flow with square rib (far-back region) - EBIV}
        \label{fig:subfig4}
    \end{subfigure}

        \begin{picture}(0,0) % Create a picture environment with zero size
        \put(-200,540){(a) Water jet flow - PIV}
        \put(10,540){(b) Water jet flow - EBIV}
        \put(-200,260){(c) Channel flow (far-back region) - PIV}
        \put(10,260){(d) Channel flow (far-back region) - EBIV}
    \end{picture}
    
    \caption{Pixel displacement computed over 150 snapshots for both experiments. The top row (a, b) shows results for the submerged water jet flow, while the bottom row (c, d) displays the channel flow with a square rib in the far-back region. The left column (a, c) corresponds to PIV measurements, and the right column (b, d) to EBIV data.}
    \label{fig:peaklock}
\end{figure*}

Figure \ref{fig:peaklock} illustrates the analysis for the presence of peak locking. In the jet flow experiment, no significant peak locking was observed (Fig.~\ref{fig:peaklock}a, Fig.~\ref{fig:peaklock}b). However, the situation is different in the squared rib case, where all velocity measurements using EBIV exhibit clear peak-locking (Fig.~\ref{fig:peaklock}d). Similarly, conventional PIV measurements still display peak locking, especially in the far-back region of the flow (Fig.~\ref{fig:peaklock}c).

Peak locking is expected in the EBIV measurements since the binary pseudo-images generated by the EBV camera limit the sub-pixel accuracy typically achievable with conventional PIV. Whereas applying a Gaussian filter can smooth out peak-locking effects for larger particles, such as those imaged in the jet flow, this method proves ineffective in the air experiment. Here, most particles appear as single-pixel activations, which prevents the filter from alleviating the issue, resulting in particle displacements being rounded to integer values.

The PIV acquisition, on the other hand, was characterized by a magnification factor that was double that of the EBV camera due to the larger pixel size. This allowed for the visualization of larger particles but required the lowest possible lens aperture ($f_\#$) to ensure sufficient particle brightness. However, this reduced the diffraction spot size of the particles (with the diffraction spot size $d_s \propto f_\#$), ultimately leading to peak-locking in the PIV measurements as well. 

As previously stated in this section, the velocity fields were interpolated onto a common grid for both cameras. Since the focus of the study is on mode extraction using POD, interpolation artifacts are expected to primarily affect the higher-order modes, without significantly impacting the analysis presented here.
The grid dimensions and total number of grid points $N_p$ obtained for the jet flow and the three square rib cases — front, back, and far-back — are presented in table \ref{tab:grid_Np}.

\begin{table}[h]
\centering
\caption{Comparison of grid dimensions and total number of grid points \( N_p \) for the considered experimental datasets.}
\renewcommand{\arraystretch}{1.5} % Increase the interline spacing
\begin{tabular}{p{2cm} cc} % Define the column types
\toprule
& Grid Dimensions & $N_p$ \\ 
\midrule
Jet Flow & $81\times152$ & 12\,312 \\ 
% \midrule
Channel Flow Front & $48\times62$ & 2\,976 \\ 
% \midrule
Channel Flow Back & $52\times93$ & 4\,836\\ 
% \midrule
Channel Flow Far-Back & $56\times83$ & 4\,648\\ 
% \midrule
% Additional Row & & \\ 
\bottomrule
\end{tabular}
\label{tab:grid_Np}
\end{table}

\subsection{Modal analysis of PIV and EBIV data}
In order to assess the performance of EBIV in identifying low-order coordinates, a modal analysis was conducted using \acf{POD} via the snapshot method by \citet{sirovich1987turbulence}. 
The snapshot matrix \( \mathbf{U} \) has dimensions \( 2N_p \times N_t \) and is formed by stacking individual velocity snapshots \( \mathbf{u}_i \) as columns:
\begin{equation}
\mathbf{U} = 
\begin{bmatrix}
\mathbf{u}_1 & \mathbf{u}_2 & \cdots & \mathbf{u}_{N_t}
\end{bmatrix},
\end{equation}
Applying \ac{POD} to this matrix yields the following decomposition:
\begin{equation}
\mathbf{U} = \mathit{\Phi} \mathit{\Sigma} \mathit{\Psi}^\top,
\end{equation}
\noindent where \( \mathit{\Phi} \) contains the spatial modes as its columns, \( \mathit{\Psi} \) contains the temporal modes as its columns, and \( \mathit{\Sigma} \) is a diagonal matrix of singular values $\sigma$. \\
The synchronized experiments enabled a comparison not only of the spatial modes, but also of the temporal modes obtained by the two measurement techniques. This result is crucial for flow control purposes, as it addresses whether EBV technology can capture the same flow dynamics as conventional PIV cameras, thus allowing real-time responses to flow changes.

A direct comparison of the temporal modes obtained from EBIV and PIV measurements is not directly possible, since the two corresponding spatial bases are not necessarily ``aligned''.\\
For this reason, a common spatial basis was established, and the velocity results were projected onto it. A training subset \(\hat{U} \in \mathbb{R}^{2N_p \times \hat{n}_t}\) of velocity fields was extracted from the PIV measurements for both experiments. These fields were randomly sampled from the first \(3 \cdot 10^3\) snapshots for the jet flow and the first \(20 \cdot 10^3\) for the channel flow. In both cases, \(\hat{n}_t\) was set to 2000.
The training set was used to create a common spatial basis from the PIV data, \(\mathit{\hat{\Phi}}_{\text{PIV}} \hat{\mathit{\Sigma}}_{\text{PIV}}\), by applying Singular Value Decomposition (SVD) to \(\hat{U}\).
Afterward, a testing dataset \(\tilde{U} \in \mathbb{R}^{2N_p \times \tilde{N}_t}\), comprising all the final \(\tilde{N}_t\) velocity snapshots, was extracted from each measurement and projected onto the common basis.\\
The computation of the spatial basis \(\mathit{\hat{\Phi}}_{\text{PIV}} \hat{\mathit{\Sigma}}_{\text{PIV}}\) can be run offline, i.e. it forms part of the training and does not necessarily require real-time processing. The temporal coefficients $\mathbf{\tilde{\psi}}_j$ of a generic snapshot $\mathbf{\tilde{u}}_j$ in the testing set, on the other hand, should be computed in real-time, following this operation:
\begin{equation}
\mathbf{\tilde{\psi}}_j = \mathbf{\tilde{u}}_j^\top \mathit{\hat{\Phi}}_{\text{PIV}} \hat{\mathit{\Sigma}}_{\text{PIV}}^{-1},
\label{eq:projpsi}
\end{equation}
In this work, however, both operations are performed offline.\\
Table \ref{tab:datasets} summarizes the details of the projections into the common basis.

\begin{table}[]
\centering
\begin{tabular}{lccc}
\toprule
{Dataset} & $N_t$ & $\hat{n}_t$ & $\tilde{N}_t$ \\%& $N_{\text{init}}$ \\
\midrule
{Jet flow} & 3\,500 & 2\,000 & 500 \\% & 3000 \\
{Channel flow} & 30\,000 & 2\,000 & 10\,000 \\% & 25000 \\
\bottomrule
\end{tabular}
\caption{Summary of training and testing dataset dimensions for jet and channel flow experiments.}
\label{tab:datasets}
\end{table}

\section{Flow statistics and spectra comparison} \label{sec:stat}
\subsection{Submerged water jet flow}
\begin{figure}[ht]
    \centering
    \includegraphics{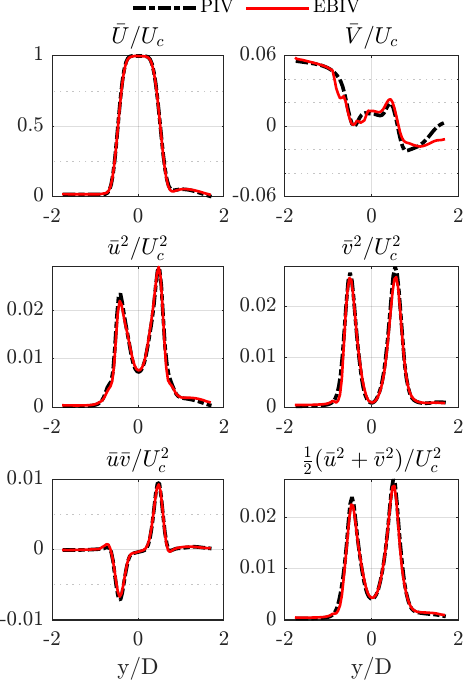}
    \caption{Statistics comparison between the conventional PIV (black dashed line) and the EBIV (red line) results. Statistics are plotted at the longitudinal location $x/D = 2$.}
    \label{fig:stats}
\end{figure}

First, the jet flow experiment is considered. Representative velocity profile statistics are presented in Fig.~\ref{fig:stats}, exhibiting consistent behavior throughout the entire domain. Velocity profiles are extracted at \( x/D = 2 \) and normalized using the jet centreline velocity \( U_c \), which is estimated as the velocity at this specific \( x/D \) location, for \( y/D = 0\). Both measurements are normalized using their respective \( U_c \). Very good agreement is observed between the two measurements.
The primary discrepancy in EBIV lies in the average radial velocity component, $\Bar{V}$. Given that the magnitude of this component is considerably smaller than that of the streamwise velocity, this discrepancy may stem from alignment uncertainties. Additionally, the more general non-symmetric behavior of $\Bar{V}$ could result from the non-canonical jet configuration: the jet is submerged but features a free upper surface, which promotes flow deflection towards the free surface.

A notable difference between the two measurements is observed in the spectral content in time of the velocity components. Fig.~\ref{fig:psd} illustrates the behaviour of the \ac{PSD} of the normalized fluctuating component of the axial velocity $u/U_b$ in the flow domain. Analogous results have been observed for the radial component $v$.
 Fig.~\ref{fig:psd}b and Fig.~\ref{fig:psd}c compare the \acp{PSD} from PIV and EBIV computed at the grid points \( (2, 0) \) and \( (5, 0.5) \), in non-dimensional coordinates, corresponding to the jet core and shear layer, respectively. These points are highlighted by the red diamonds in Fig.~\ref{fig:psd}a.
Good agreement is observed in the frequency range $\mathit{St} < 1.5$, where the Strouhal number $\mathit{St}$ is obtained normalizing the frequencies by the nozzle diameter $D$ and the jet bulk velocity $U_b$. At both points, EBIV exhibits different behavior for frequencies associated with $\mathit{St} > 1.5$, but with opposite trends in the two regions of the jet. In the core, the energy is significantly higher at higher frequencies, whereas within the shear layer, it is lower. 
% \todo{Normally the flow at the jet's core is laminar up to $x/D \approx 5$ along the centerline. Have you checked this? What is surprising though, is that EBIV seems to dampen the higher frequencies}
This difference can be attributed to the higher noise level in the pseudo-images generated from the EBV camera data stream. The noise, not tied to any specific frequency, is distributed across the entire spectrum, resulting in the core in a higher plateau reflecting its intensity. This occurs because the core area is primarily dominated by the shedding frequency of the jet vortex rings. In contrast, the shear layer is characterized by a broadband turbulent spectrum. The different behavior in the shear layer is ascribed to a higher number of outliers detected during the EBIV processing and their subsequent removal, leading to lower energy content. The filtering out of the higher frequency content can be ascribed to the higher measurement uncertainty. 

The maps displayed in plots d, e, f and g of Fig.~\ref{fig:psd} are generated by calculating the \ac{PSD} at each grid point along lines parallel to the $x$-direction, specifically at $y/D = 0$ (Fig.~\ref{fig:psd}d, Fig.~\ref{fig:psd}e) and $y/D = 0.5$ (Fig.~\ref{fig:psd}f, Fig.~\ref{fig:psd}g). These lines are depicted on the velocity map of Fig.~\ref{fig:psd}a with black and cyan dashed lines, respectively, and the maps are outlined with corresponding color-coded frames. Results from PIV are shown on the left (Fig.~\ref{fig:psd}d, Fig.~\ref{fig:psd}f), while those from EBIV are presented on the right (Fig.~\ref{fig:psd}e, Fig.~\ref{fig:psd}g).
% related to a specific line have frames colored to match the corresponding line color.
\definecolor{customcolor}{rgb}{0,1,1}
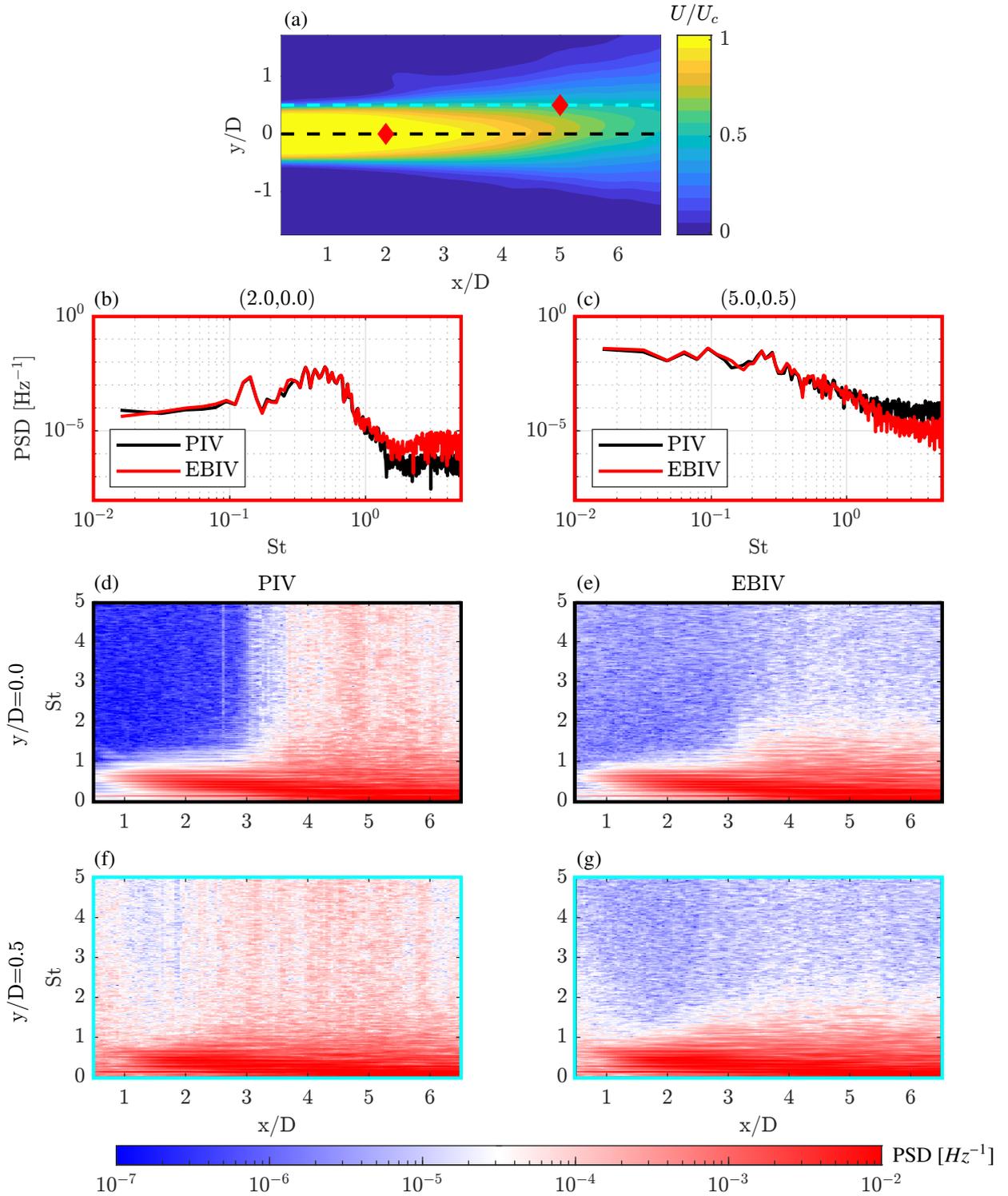
\begin{figure*}[]
    \centering
    \begin{tikzpicture}[remember picture]
        % Include the image
        \node[anchor=south west,inner sep=0] (image) at (0,0) {\includegraphics{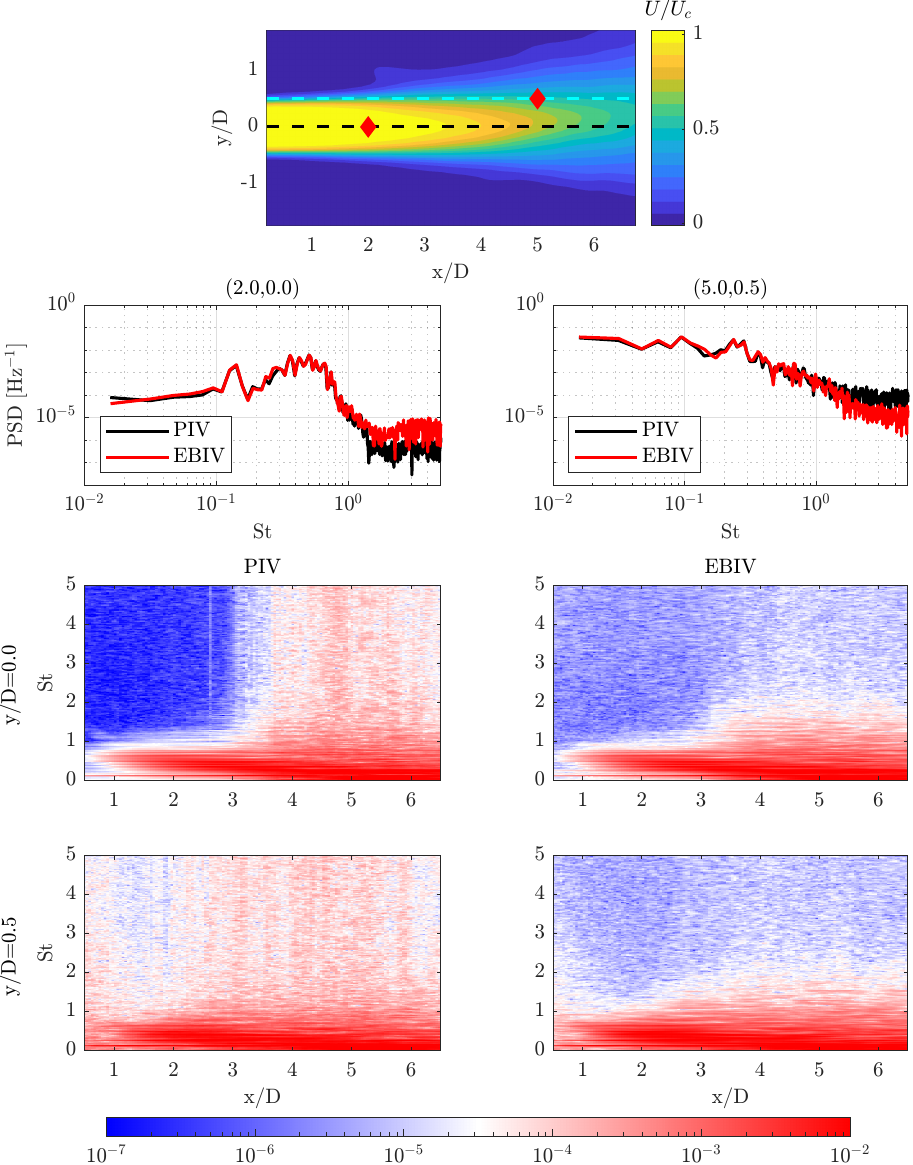}};
        % Define the coordinates for the rectangle (choose your desired position)
        \coordinate (top_left) at (1.43,5.25);
        \coordinate (bottom_right) at (7.48,1.92);
        % Draw the rectangle
        \draw[ultra thick, customcolor] (top_left) rectangle (bottom_right);

        % Define the coordinates for the second rectangle
        \coordinate (top_left2) at (9.35,5.25);
        \coordinate (bottom_right2) at (15.4,1.92);
        % Draw the second rectangle
        \draw[ultra thick, customcolor] (top_left2) rectangle (bottom_right2);
        
        % Define the coordinates for the third rectangle (top left)
        \coordinate (top_left3) at (1.43,9.8);
        \coordinate (bottom_right3) at (7.48,6.5);
        % Draw the second rectangle
        \draw[ultra thick, black] (top_left3) rectangle (bottom_right3);

        % Define the coordinates for the forth rectangle
        \coordinate (top_left4) at (9.35,9.8);
        \coordinate (bottom_right4) at (15.4,6.5);
        % Draw the second rectangle
        \draw[ultra thick, black] (top_left4) rectangle (bottom_right4);

        % Define the coordinates for the fifth rectangle (top left)
        \coordinate (top_left5) at (1.43,14.55);
        \coordinate (bottom_right5) at (7.48,11.5);
        % Draw the second rectangle
        \draw[ultra thick, red] (top_left5) rectangle (bottom_right5);

        % Define the coordinates for the forth rectangle
        \coordinate (top_left6) at (9.35,14.55);
        \coordinate (bottom_right6) at (15.4,11.5);
        % Draw the second rectangle
        \draw[ultra thick, red] (top_left6) rectangle (bottom_right6);
    \end{tikzpicture}

    \begin{picture}(0,0) % Create a picture environment with zero size
        \put(-90,563){(a)}
        \put(-179,431){(b)}
        \put(47,431){(c)}
        \put(-179,297){(d)}
        \put(47,297){(e)}
        \put(-179,167){(f)}
        \put(47,167){(g)}
        \put(195,27){PSD [\( Hz^{-1}\)]}
    \end{picture}
    
    \caption{\Acf{PSD} of the normalized $u/U_b$ velocity component of the jet flow. The red box compares core (b) and shear layer (c) measurements. Locations are marked by red diamonds in the top plot. The black (d,e) and cyan (f,g) boxes represent \ac{PSD} maps along $y/D=0$ and $y/D=0.5$, respectively. Left (d,f) shows PIV results, while right (e,g) shows EBIV results.}
    \label{fig:psd}
\end{figure*}

% \begin{figure}
%     \centering
%     \includesvg{Fig/SpectraFullSVG.svg}
%     \caption{Power Spectral Density (PSD) of the $u$ velocity component of the jet flow. The red box highlights the comparison between the two measurements in the core (b) and in the shear layer (c). The locations are indicated in the top plot with red diamonds and the coordinates are shown on top of the plots. The black (d,e) and cyan (f,g) boxes represent the PSD maps along the lines $y/D=0$ and $y/D=0.5$, respectively, as shown in the top plot. On the left (d,f) are the results from PIV, while on the right (e,g) are those from EBIV.}
%     \label{fig:psd}
% \end{figure}

In general, the characteristics observed in the single-point \ac{PSD} hold for the entire flow domain. The most energetic frequencies are accurately described by EBIV, both in the shear and core regions. However, for $\mathit{St} > 1.5$, corresponding to frequencies $f > 10\,\text{Hz}$, significant differences arise: higher \ac{PSD} values are observed in the core, while attenuation at higher frequencies occurs in region of high turbulence. The latter trend is evident throughout the maps related to the line $y/D = 0.5$, and in the case of $y/D = 0$ for $x/D \geq 4$, corresponding to the core collapse.

\subsection{Channel air flow with squared rib}
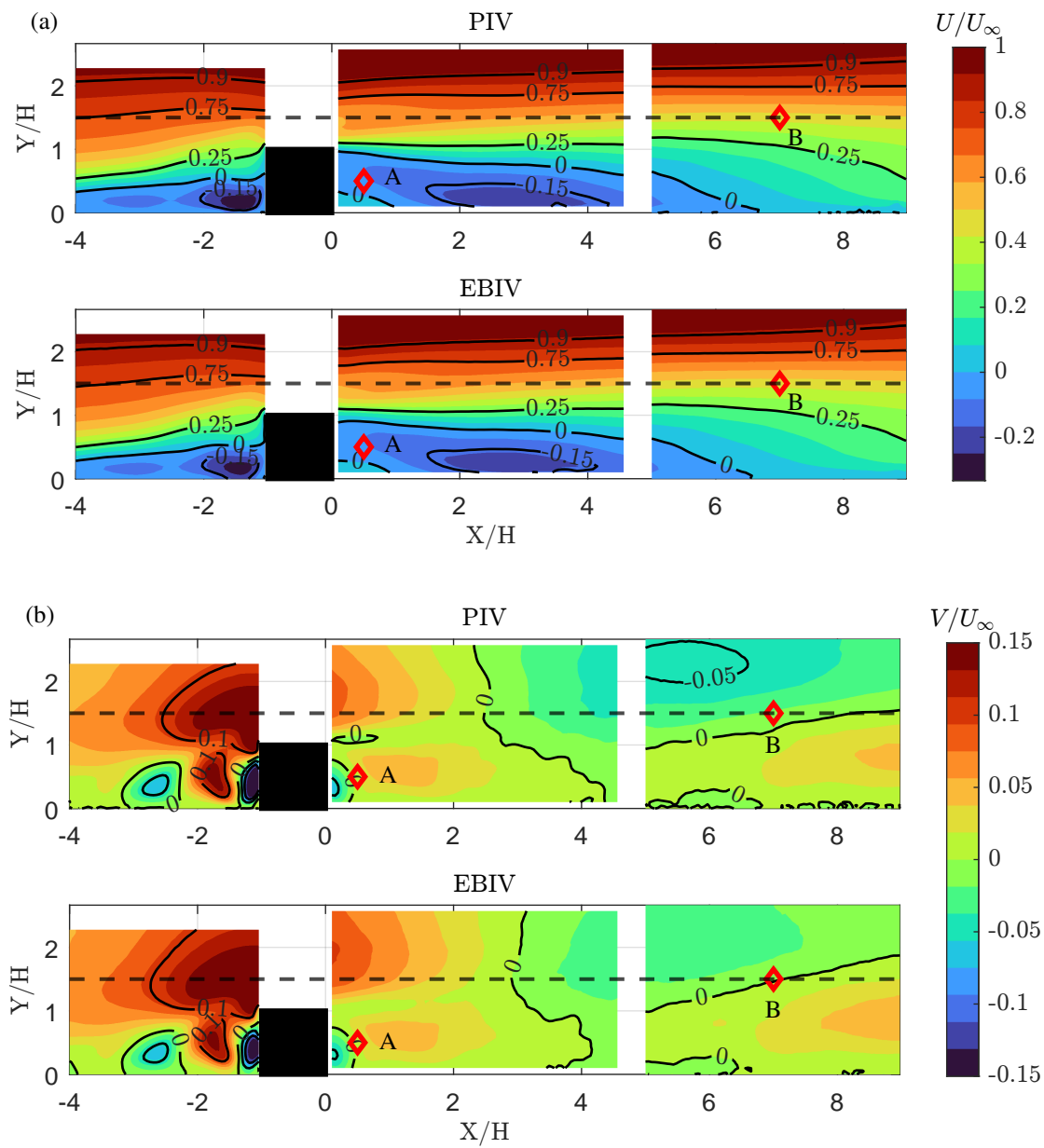
\begin{figure*}[htbp]
    \centering
    \begin{subfigure}[b]{\textwidth}
        \centering
        \includegraphics{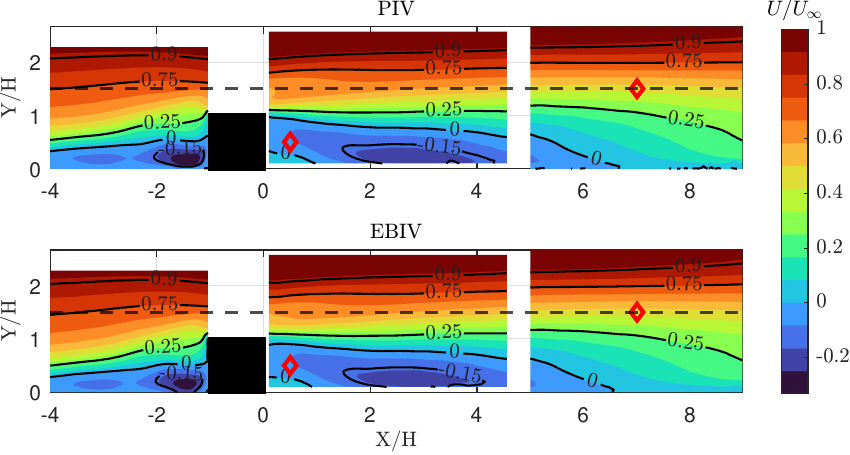}
        % \caption{}
        \label{fig:image1}
    \end{subfigure}
    
    \vspace{0.75cm} % Add some vertical space between the images

    \begin{subfigure}[b]{\textwidth}
        \centering
        \includegraphics{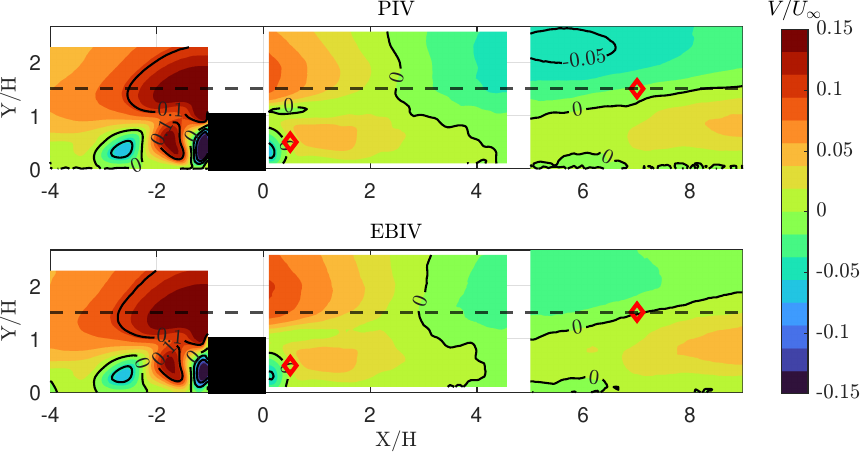}
        % \caption{}
        \label{fig:image2}
    \end{subfigure}
    
    \begin{picture}(0,0) % Create a picture environment with zero size
        \put(-200,465){(a)} % Adjust coordinates to position (a) near the first figure
        \put(-202,225){(b)} % Adjust coordinates to position (b) near the second figure
        \put(-58,402){A}
        \put(104,418){B}
        \put(-58,294){A}
        \put(104,311){B}

        \put(-60,160){A}
        \put(95,172){B}
        \put(-60,52){A}
        \put(95,65){B}
    \end{picture}
    \caption{Average flow velocities $U$ (a) and $V$ (b) in the $X$ and $Y$ directions, respectively, for the channel flow experiment, normalized by the free-stream velocity $U_\infty$. In both figures, the top row presents the PIV results, while the bottom row shows the EBIV results. Velocity isolines corresponding to specific velocity values are displayed. The black horizontal line indicates the $Y/H$ coordinate for which the second-order statistics are presented in Fig.~\ref{fig:stat2sr}. The red diamonds mark the locations of the \ac{PSD} plots in Fig.~\ref{fig:srpsd}.
    }
    \label{fig:srstat}
\end{figure*}

Figure \ref{fig:srstat} presents the average velocities in the $X$ (Fig.~\ref{fig:srstat}a) and $Y$ (Fig.~\ref{fig:srstat}b) directions, $U$ and $V$, for both PIV (top) and EBIV (bottom), normalized by the free-stream velocity $U_\infty$. The figure also includes isolines with specific velocity values labeled directly on the lines.
Overall, EBIV reliably captures the flow statistics, particularly the reattachment point of the wake, which occurs approximately 7$H$ downstream of the rib, as well as the recirculation areas both in front of and behind the obstacle. The $V$ component shows more noticeable differences, especially in the 0-velocity isolines. However, given the small magnitudes involved, these discrepancies can likely be attributed to alignment issues. 

\begin{figure*}
    \centering
    \includegraphics{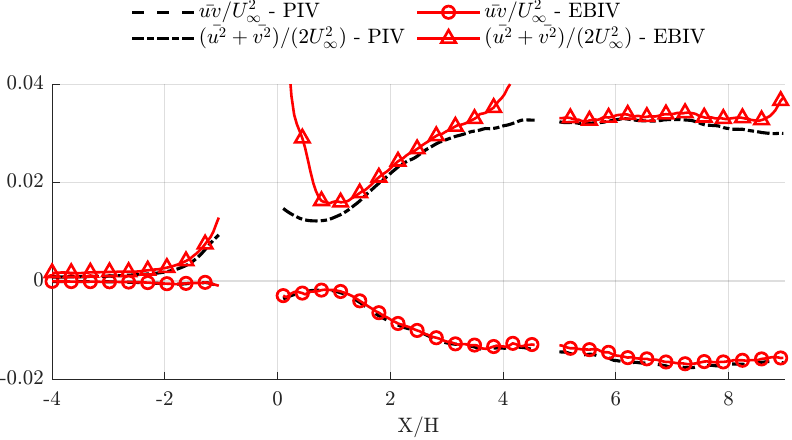}
    \caption{Average normalized second order flow statistics for the channel flow experiment, along the horizontal line $Y/H = 1.5$. In black, the reference values from PIV. In red, the results from EBIV. For sake of representation, only one marker every 7 is shown.}
    \label{fig:stat2sr}
\end{figure*}

Figure \ref{fig:stat2sr} shows part of the normalized second-order statistics. Specifically, the Reynolds shear stress, $\Bar{uv}/U_\infty^2$, and the \ac{TKE}, $(\Bar{u^2} + \Bar{v^2})/(2 U_\infty^2)$, are compared along the dashed line at ${Y/H} = 1.5$, as shown in Fig.~\ref{fig:srstat}. PIV results, considered as the reference, are represented by black lines, while EBIV measurements are plotted using red markers with a continuous line. Similar results can be observed throughout the flow domain.

Good agreement is observed across most of the streamwise direction. However, notable discrepancies appear at the edges of the \ac{FOV} for the \ac{TKE}, particularly in the region behind the rib. The instantaneous velocity snapshots contain outliers due to a lack of particles in the EBIV images. While this issue can be mitigated by applying multi-frame cross-correlation (CC) processing, simpler two-time-step CC was considered for the purposes of this study. The main argument is that we aim to compare the capability to extract \acp{ROM} from EBIV and PIV data assuming the most slender processing possible for EBV images, in view of future online implementation. The particle deficit is not due to insufficient laser intensity but rather the large number of events processed by the EBV sensor and the non-uniformity of the laser intensity across the sheet. Operating near its event-processing capacity, the EBV sensor primarily captures the most intense particle scatter. With the laser sheet having a Gaussian intensity distribution across the horizontal \ac{FOV}, most particles are captured at the center of the \ac{FOV}. The edges of the \ac{FOV} suffer from a lack of particles, leading to increased velocity fluctuations ($u$ and $v$). While these fluctuations were filtered out in the averaged velocities shown in Fig.~\ref{fig:srstat}, they become apparent in the second-order statistics. This issue was partially mitigated in other measurement regions by further expanding the laser sheet to achieve a more uniform light intensity across the plane, resulting in better agreement.

\begin{figure}
    \centering
    \includegraphics{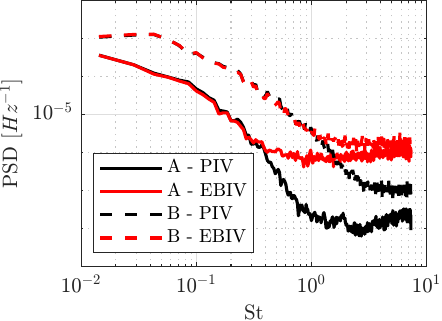}
    \caption{\ac{PSD} of the normalized $u/U_\infty$-velocity fluctuation in the points A = ($0.5$, $0.5$) and B = ($7$, $1.5$), indicated by red diamonds in Fig.~\ref{fig:srstat}. The frequencies are normalized as $\textit{St} = f\frac{H}{U_\infty}$.}
    \label{fig:srpsd}
\end{figure}

The \ac{PSD} of the normalized x-velocity component $u/U_\infty$ is compared between PIV and EBIV in Fig.~\ref{fig:srpsd}. The \ac{PSD} is calculated at two sample points, defined in terms of $X/H$ and $Y/H$: point A = ($0.5$, $0.5$) and point B = ($7$, $1.5$), as indicated by the red diamonds in Fig.~\ref{fig:srstat}. Frequencies are normalized to Strouhal numbers defined as  $\mathit{St} = f\frac{H}{U_\infty}$. No specific frequency signature can be observed in the flow field. A peak in the velocity spectra was expected due to vortex shedding behind the rib. However, the flow exhibited significant three-dimensionality, likely caused by the mounting of the obstacle in the channel, which generated side effects. It is probable that the out-of-plane motion suppressed the typical vortex shedding signature \citep{rashidi2016vortex}, resulting in a more uniformly distributed turbulent spectrum.

In general, EBIV exhibits a significantly higher noise level compared to PIV, resulting in an elevated plateau of the \ac{PSD} at higher frequencies. This plateau is found to be of similar magnitude for both the $u$ and $v$ PSDs and is primarily dependent on the location of the sample point. Overall, this finding is consistent across the full flow domain and demonstrates that EBIV can effectively capture the most energetic frequencies of the flow, albeit above a noise floor that is several orders of magnitude higher than that of PIV.

\section{Modal analysis}

% \begin{comment}{
% A modal analysis was conducted using \acf{POD} via the snapshot method by \citet{sirovich1987turbulence}. The snapshot matrix \( \mathbf{U} \) has dimensions \( 2N_p \times N_t \) and is formed by stacking individual velocity snapshots \( \mathbf{u}_i \) as columns:
% \begin{equation}
% \mathbf{U} = 
% \begin{bmatrix}
% \mathbf{u}_1 & \mathbf{u}_2 & \cdots & \mathbf{u}_{N_t}
% \end{bmatrix},
% \end{equation}
% Applying \ac{POD} to this matrix yields the following decomposition:
% \begin{equation}
% \mathbf{U} = \mathit{\Phi} \mathit{\Sigma} \mathit{\Psi}^\top,
% \end{equation}
% \noindent where \( \mathit{\Phi} \) contains the spatial modes as its columns, \( \mathit{\Psi} \) contains the temporal modes as its columns, and \( \mathit{\Sigma} \) is a diagonal matrix of singular values $\sigma$. These computations were performed across the entire set of available snapshots from both the PIV and EBIV experiments.} 
% \end{comment}

\begin{figure*}[]
    \centering
    % First subfigure
    \begin{subfigure}[]{\textwidth}
        \centering
        \includegraphics{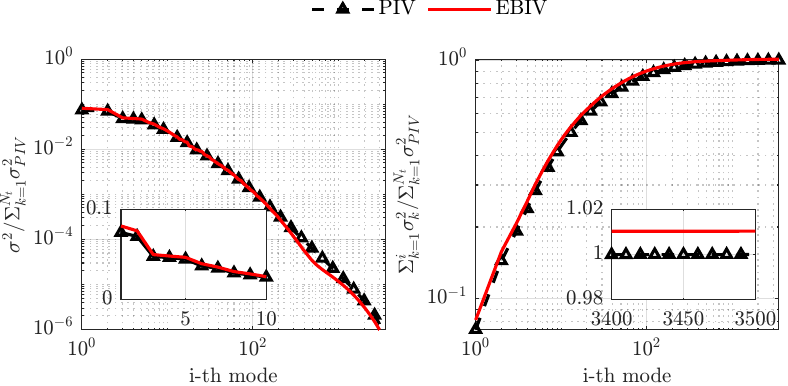}
        % \caption{Jet flow experiment}
        \label{fig:sigmajet}
    \end{subfigure}
    % \hfill % Adds space between the two figures

    \vspace{0.5cm}

    % Second subfigure
    \begin{subfigure}[]{\textwidth}
        \centering
        \includegraphics{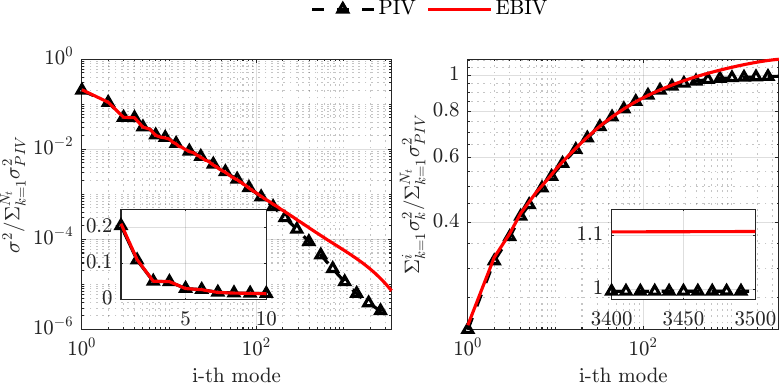}
        % \caption{Channel flow with square rib experiment}
        \label{fig:sigmasr}
    \end{subfigure}

    \begin{picture}(0,0) % Create a picture environment with zero size
        \put(-190,382){(a)} % Adjust coordinates to position (a) near the first figure
        \put(-190,184){(b)} % Adjust coordinates to position (b) near the second figure
    \end{picture}
    
    \caption{Mode energy comparison for two experimental datasets: (a) jet flow and (b) channel flow with a square rib, far-back region. Each plot presents the energy distribution per mode (left) and the cumulative energy (right), both normalized by the total energy from PIV measurements. Black triangular markers represent PIV results, while red lines correspond to EBIV data. For clarity, only the first \(10^3\) modes are shown on the left plot and the first \(3.5 \times 10^3\) modes on the right. Both plots use logarithmic scales on both axes. Markers are distributed logarithmically across the plot to better represent the variation in the data. A linear zoom-in is provided for better visualization: in the single-mode energy plot, the first 10 modes are emphasized, while in the cumulative energy plot, modes between 3400 and 3500 are enlarged.}
    \label{fig:enmodes} 
\end{figure*}

A comparison of the energy distribution of the modes, derived from the squared singular values \( \sigma \), is presented in Fig.~\ref{fig:enmodes}. For both the jet flow (a) and the far-back region of the channel flow (b), a comparison of the squared singular values (left) and the cumulative sum of \( \sigma_k^2 \) (right) is shown. The reference from PIV is indicated in black, while the EBIV results are shown in red. Each sequence is normalized by the cumulative sum of squared singular values from the PIV measurements.

In both cases, EBIV correctly identifies the significance of the dominant modes. However, the cumulative energy plot reveals that EBIV measures higher energy than PIV. In the jet flow experiment, the overshoot in energy is limited to 1\% compared to PIV. In contrast, in the channel flow experiment—especially in the back region—the overshoot reaches up to 25\,\%.

In the jet flow case, EBIV slightly overestimates the energy of the first few modes, but the results show that the energy of higher-order modes is slightly lower than that of PIV, compensating for the initial discrepancy and leading to a minor overall overestimation of turbulent kinetic energy compared to PIV. This discrepancy, around 1\%, is likely due to a misalignment between the two flow fields rather than an inherent limitation of the measurement technique, as it is not consistently observed across other datasets.

For the channel flow, the first and most energetic modes are well reconstructed by EBIV. However, higher-order modes exhibit greater contamination by noise, resulting in a cumulative energy sum that is 11\% higher than that of PIV. The most significant deviation is observed in the wake region behind the rib, where EBIV overestimates the total energy by up to 25\%, largely due to the presence of outliers at the edges of the domain, as discussed in section \ref{sec:stat}.\\

\begin{figure*}[]
    \centering
    \begin{subfigure}[]{\textwidth}
        \centering
        \includegraphics{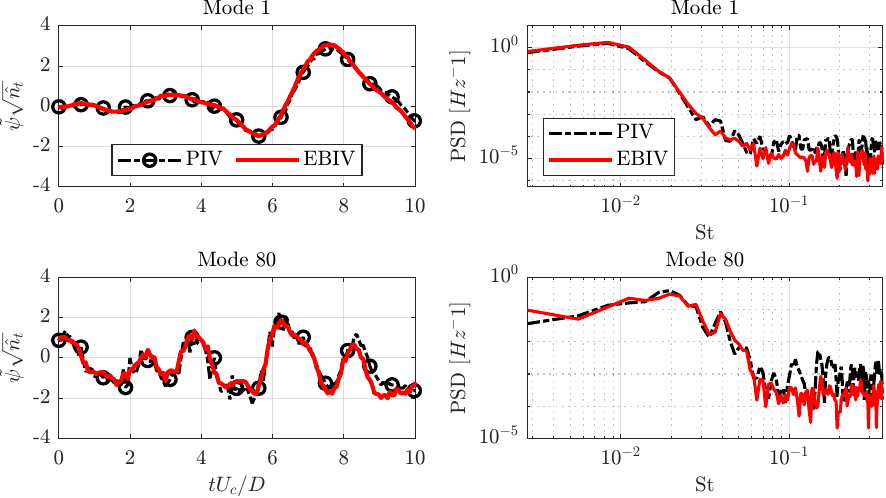}
        % \caption{}
        \label{fig:psijet}
    \end{subfigure}

    \vspace{0.75cm}
    
    \begin{subfigure}[]{\textwidth}
        \centering
        \includegraphics{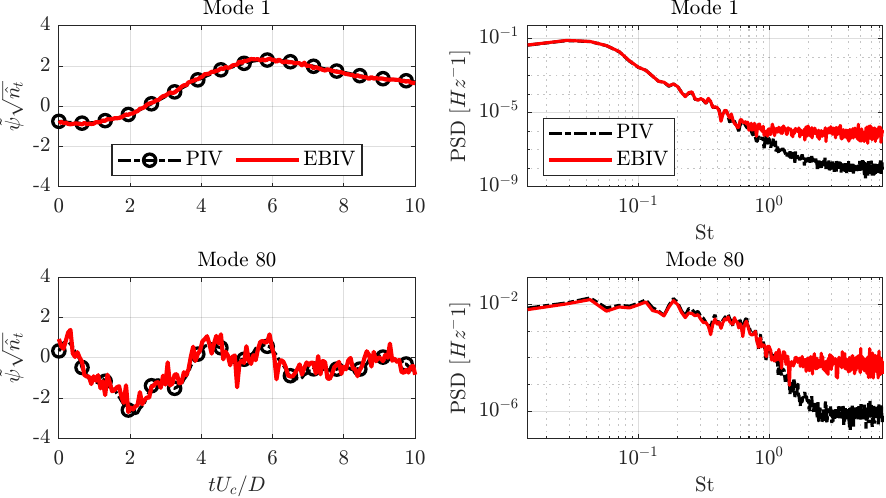}
        % \caption{}
        \label{fig:psisr}
    \end{subfigure}

    \begin{picture}(0,0) % Create a picture environment with zero size
        \put(-210,512){(a)} % Adjust coordinates to position (a) near the first figure
        \put(-210,245){(b)} % Adjust coordinates to position (b) near the second figure
    \end{picture}

    \caption{On the left, the projected temporal mode coefficients $\tilde{\psi}$ for modes 1 (top) and 80 (bottom) are shown over a duration of 10 convective time units. These modes were obtained by projecting the fluctuating velocity fields onto a common spatial basis \(\mathit{\hat{\Phi}}_{\text{PIV}} \hat{\mathit{\Sigma}}_{\text{PIV}}\), derived from the \ac{PIV} measurements. The jet flow is shown in (a), and the far-back region of the channel flow in (b). The dashed black line with markers represents the results from PIV, while the red line represents the EBIV results. The values are normalized by the square root of the training dataset's dimension, $\hat{n_t}$, used to compute the common basis. Only one marker out of every ten is displayed. On the right, the \ac{PSD} of the two projected temporal modes is color-coded according to the measurement technique.}
    \label{fig:psiproj}
\end{figure*}

The comparisons between the projected temporal mode coefficients \(\mathbf{\tilde{\psi}_j}\), with respect to a common spatial basis,\(\mathit{\hat{\Phi}}_{\text{PIV}} \hat{\mathit{\Sigma}}_{\text{PIV}}\), as described in Eq.~\ref{eq:projpsi}, are presented in Fig. ~\ref{fig:psiproj}. Modes 1 and 80, corresponding to the jet flow (Fig.~\ref{fig:psiproj}a) and the channel flow (Fig.~\ref{fig:psiproj}b) are shown. On the left, the values of the projected temporal mode coefficients \(\tilde{\psi}\) are plotted, normalized by the square root of the number of snapshots \( \hat{n_t} \) in the training dataset used to compute the common spatial basis. The reference value from PIV is shown in black, while the results from \ac{EBIV} are depicted in red. The projected temporal modes are plotted for a short sequence lasting 10 convective units within the testing set.
On the right, the \ac{PSD} of the aforementioned projected temporal modes \(\tilde{\Psi}\) is shown, maintaining the same color coding as before.\\
Even for higher mode numbers, which are typically subject to noise, \ac{EBIV} demonstrates its ability to accurately capture the same dynamics as \ac{PIV}. In the channel flow case, the temporal modes from EBIV exhibit more fluctuations but generally follow the primary trends of the mode. This behavior is consistent across the entire flow domain.
Moreover, the \ac{PSD} plot confirms the same trend observed in the \ac{PSD} of the fluctuating velocity fields. In the jet flow, the high-frequency plateau is higher for the PIV measurements, as shown in Fig.~\ref{fig:psd}, where the \ac{PSD} in the mixing layer—energetically dominant in the field—exhibits the same behavior. The low-frequency component is well reconstructed by \ac{EBIV}, even for relatively high mode numbers.\\
A different behavior is observed in the channel flow experiment. Low-order modes accurately capture the high-energy low-frequency content but plateau at a higher level at higher frequencies. For higher-order modes, an offset is noticeable at lower frequencies, while the noise plateau approaches the main frequency energy level, indicating that noise becomes increasingly dominant.
Overall, across both measurement techniques and flow conditions, the high-frequency plateau increases with higher mode numbers, confirming the expected correlation between mode number and noise susceptibility.\\

More generally, assuming the PIV results as the ground truth, the normalized root mean square error (RMSE) between the $i$-th EBIV and PIV projected temporal coefficients \( \Tilde{\psi}_{i} \), scaled by the $i$-th common singular value \( \hat{\sigma}_{\text{PIV},i}\), can be introduced. This error is defined for the $\tilde{N}_t$ testing subset as follows:

\begin{equation}     
   \epsilon_{\hat{\sigma}\Tilde{\Psi}^T}(i) = \sqrt{ \frac{\hat{n}_t}{\tilde{N}_t}      \frac{\sum\limits_{j=1}^{\tilde{N}_t} \hat{\sigma}_{\text{PIV},i}^2 \left( \Tilde{\psi}_{\text{PIV},i}(j) - \Tilde{\psi}_{\text{EBIV},i}(j) \right)^2}{\sum\limits_{k=1}^{\hat{n}_t} \hat{\sigma}_{\text{PIV},k}^2}}
    \label{eq:error_psi}
\end{equation}

Note that the denominator in Eq. \ref{eq:error_psi} is related to the normalization of the error.
\begin{figure*}[]
    \centering
    \includegraphics{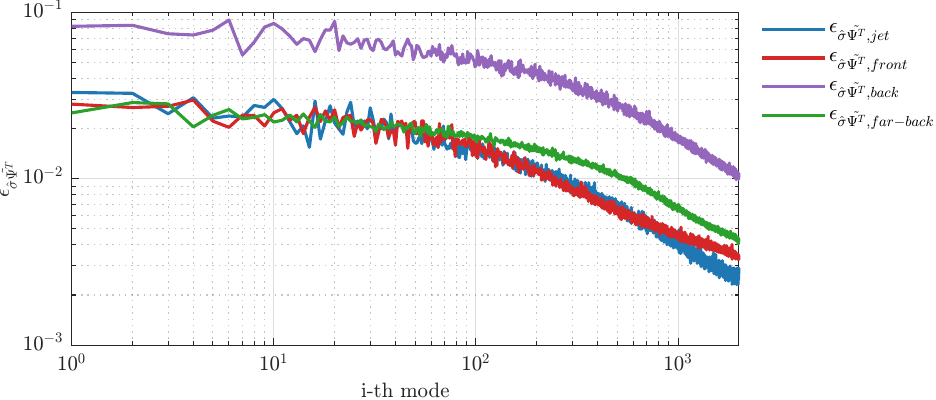}
    \caption{Root Mean Square Error (RMSE) of the \(i\)-th projected temporal modes \( \Tilde{\Psi}_{i} \) between PIV and EBIV, scaled by the corresponding singular value \( \hat{\sigma}_{\text{PIV},i} \). The blue line represents the RMSE for the jet flow experiment, while the red, purple, and green lines correspond to the front, back, and far-back regions of the channel flow, respectively. The entire set of \( \hat{n}_t \) modes is presented.}
    \label{fig:errmode}
\end{figure*}

The projection errors, calculated only for the testing dataset, are shown in Fig.~\ref{fig:errmode} for both the jet flow (blue) and the three regions of the channel flow experiments: front (red), back (purple), and far-back (green). 

Lower-order modes, despite exhibiting small errors in their shape, contribute significantly to a higher \( \epsilon_{\hat{\sigma}\Tilde{\Psi}^T} \) because they have higher energy. In contrast, higher-order modes have the opposite behaviour. As the weight of these modes decreases more rapidly than the error in \( \Tilde{\Psi} \), the overall \( \epsilon_{\hat{\sigma}\Tilde{\Psi}^T} \) gradually decreases with increasing mode order.\\
Overall, the error remains within a few percentage points, with the back region of the channel flow being the most affected, showing an initial error of approximately 8.5\%. Notably, the other regions of the channel flow exhibit errors in the same order as those observed in the jet flow.\\

\begin{figure*}
    \centering
    \begin{overpic}{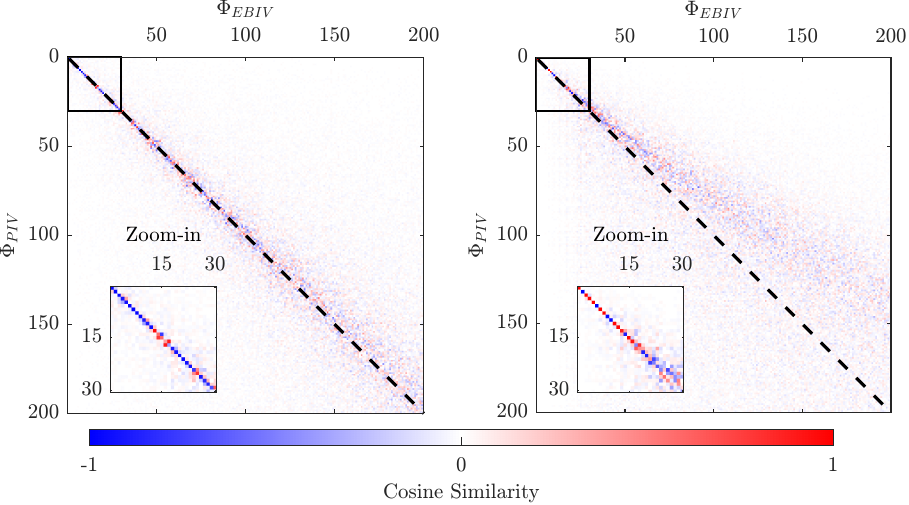}   %[grid,tics=20]
        \put(15,230){(a)}
        \put(240,230){(b)}
    \end{overpic}
    \caption{Cosine similarity matrix $CS$ between the spatial modes $\Phi$ from PIV and EBIV, in the case of the jet flow (a) and the back-rib region of the channel flow (b). Values of 1 indicates full similarity, 0 orthogonality of the modes and -1 phase opposition. Only the first 200 modes are shown. The area squared in black represents the zoom-in window for the first only 30 modes. The dashed line indicates the matrix diagonal.}
    \label{fig:cossim}
\end{figure*}

The accuracy of spatial mode reconstruction is evaluated using cosine similarity. In this case, no projection onto a common basis is performed. Instead, the spatial modes $\Phi$ are directly compared after being computed via \ac{SVD}, on the full dataset $N_t$. The comparison is made between the $i$-th spatial mode from PIV and the $j$-th spatial mode from EBIV. Mathematically, the cosine similarity matrix $CS \in \mathbb{R}^{2N_p \times 2N_p}$ is computed as the scalar product between the spatial modes of the two measurements:
\begin{equation}
    CS = \Phi_{PIV} \cdot \Phi_{EBIV}^T
\end{equation}
No normalization is required, as the spatial modes already possess a unitary norm by definition. Ideally, if all spatial modes were identical, the cosine similarity matrix would display a diagonal of ones (red in the map). However, due to increasing noise contamination and potential phase opposition, the cosine similarity values tend to decrease, become negative (blue in the map), or cause highly similar modes to diverge from the diagonal. Additionally, cross-talk between neighboring modes results in blurred similarity values around the diagonal. Figure~\ref{fig:cossim} presents the cosine similarity matrices for the jet flow (Fig.~\ref{fig:cossim}a) and the back region of the squared rib measurements (Fig.~\ref{fig:cossim}b). The first 200 modes per case are displayed, with a zoom-in area focusing on the first 30 modes. 

In both cases, the main diagonal gradually fades, indicating a spreading of corresponding modes into adjacent ones for the other measurement technique. Overall, the jet flow dataset exhibits better behavior, with the fading occurring consistently along the main diagonal. In contrast, the higher noise contamination in the channel flow causes the $i$-th PIV mode to spread into higher-rank EBIV modes. This pattern is observed across all investigated regions. Despite this, very good correlation is still seen for the most energetic modes.

Overall, EBIV accurately captures both the main spatial and temporal characteristics, demonstrating robustness to noise in the raw data and producing results comparable to PIV even at higher mode numbers. For the jet flow, the primary discrepancy lies in estimating the singular values for the most energetic modes (see Fig.~\ref{fig:enmodes}), exhibiting an error of approximately 10\%. \\ % compared to the PIV values. In the case of the squared rib, the estimation of the most relevant singular value modes is accurate; however, the measurements are affected by greater noise corruption, primarily confined to the higher-order modes. 

\begin{figure*}[]
    \centering
    \includegraphics{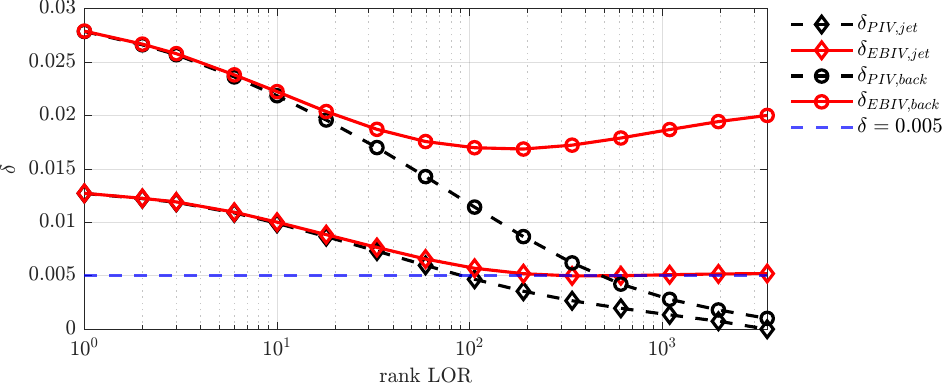}
    \caption{\ac{LOR} error \(\delta\) for both experiments, referenced to the full-rank velocity field computed via PIV, as function of the truncation rank $r$. The experiments are represented with consistent marker types: diamonds for jet flow and circles for channel flow. The measurement techniques are color-coded, with black representing PIV and red representing EBIV. For visualization purposes, only the first 3500 rank approximation errors are plotted, with markers equally distributed on a logarithmic scale.}
    \label{fig:errLOR}
\end{figure*}

Finally, a \acf{LOR} of the velocity fields \(\tilde{\mathbf{u}}_{\text{LOR},r} \in \mathbb{R}^{2N_p \times N_t}\) is obtained by reconstructing the flow using only the first \( r \) modes from the Proper Orthogonal Decomposition (POD) for both PIV and EBIV data. The reduced-order approximation of the velocity field is given by
\begin{equation}
    \tilde{\mathbf{u}}_{\text{LOR},r}(x,t) = \sum_{i=1}^r \Psi_i (t) \sigma_i \mathbf{\Phi_i}(x)
\end{equation}
The reconstruction is performed for increasing ranks \( r \), and the resulting \acp{LOR} for PIV and EBIV are compared to the full-rank PIV velocity field \(\mathbf{u}_{\text{PIV}}(x,t)\). This comparison is aimed at evaluating the dimensionality reduction performance of EBIV.\\
To quantify this evaluation, an error metric \( \delta(r) \) is introduced. This metric measures the time-averaged deviation of the reduced-order approximation \( \tilde{\mathbf{u}}_{\text{LOR},r} \) from the full-rank PIV velocity field \(\mathbf{u}_{\text{PIV}}\). For each rank \( r \), the point-wise root mean square error (RMSE) is computed for all velocity components across the entire flow field and averaged over the time series. The error is normalized by the maximum turbulent kinetic energy (TKE) observed in the PIV measurements, resulting in the following expression for \( \delta(r) \):
\begin{equation}
        \delta(r) = \frac{1}{TKE_{\text{max}}} \frac{\left\langle \left\| \mathbf{u}_{\text{PIV}}(t) - \tilde{\mathbf{u}}_{\text{LOR},r}(t) \right\|_2 \right\rangle_{t}}{\sqrt{2N_p}} 
\end{equation}
where \( \left\| \cdot \right\|_2 \) represents the \(L^2\)-norm over the spatial domain 
% \begin{equation}
%         \left\| \mathbf{u}_{\text{PIV}}(t) - \tilde{\mathbf{u}}_{\text{LOR},r}(t) \right\|_{2} = 
%         \sqrt{\sum_{j=1}^{2N_p} \left( \mathbf{u}_{\text{PIV}}(j,t) - \tilde{\mathbf{u}}_{\text{LOR},r}(j,t) \right)^2}
% \end{equation}
%In this formulation, 
and the angle brackets \(\langle \cdot \rangle\) denote time-averaging over the acquired time series. %Note that j spans over $2N_p$ since it consider both the $X$ and $Y$ velocity components.
% Finally, a \acf{LOR} of the velocity fields \(\tilde{\mathbf{u}}_{\text{LOR},r} \in \mathbb{R}^{2N_p \times N_t}\) is computed by considering only the first \( r \) modes from the Proper Orthogonal Decomposition (POD) for both PIV and EBIV, i.e.
% \begin{equation}
%     \tilde{\mathbf{u}}_{\text{LOR},r}(x,t)=\sum_{i=1}^r \Psi_i (t) \sigma_i \mathbf{\Phi_i}(x).
% \end{equation}
% The reconstruction is performed for increasing ranks \( r \), and the resulting \acp{LOR} for PIV and EBIV are directly compared to the full-rank representation from the PIV measurement, \(\mathbf{u}_{\text{PIV}}\). This comparison evaluates the dimensionality reduction capabilities of EBIV.\\
% To facilitate this evaluation, a specific error metric \( \delta(r) \) is defined. This metric quantifies the average temporal error of the \ac{LOR} \( \tilde{\mathbf{u}}_{\text{LOR},r} \) for both measurement techniques in relation to the corresponding full-rank PIV velocity snapshot \( \mathbf{u}_{\text{PIV}} \). For each rank value \( r \), the root mean square error (RMSE) of the point-wise differences across each snapshot is averaged over the acquired time series. The error is normalized by the maximum value of the mean turbulent kinetic energy measured with PIV, resulting in the following error definition:
% \begin{equation}
%     \delta(r) = \frac{1}{TKE_{\text{max}}} \left\langle  \sqrt{\frac{\sum\limits_{j=1}^{2N_p} \left( \mathbf{u}_{\text{PIV}}(j,t) - \tilde{\mathbf{u}}_{\text{LOR},r}(j,t) \right)^2}{2N_p}} \right\rangle 
% \end{equation}
The obtained values for the jet flow experiment and the back region of the squared rib in the channel flow are shown in Fig.~\ref{fig:errLOR}.\\
The error from the PIV measurement tends to zero, since including more and more modes in the \ac{LOR}, the low-rank approximation approaches the full-field reconstruction, considered as reference. For the EBIV measurements, the error reaches a minimum value of approximately 0.5\% in the jet flow case as the rank increases. In contrast, for the channel flow, an optimal rank that minimizes the error is observed between \( \text{r} = 100 \) and \( \text{r} = 200 \). At this range, the error is comparable to that of the jet flow but remains a few times higher. Beyond this rank, the error \(\delta_{\text{EBIV}}\) begins to increase and subsequently stabilizes.\\

% \begin{figure*}[]
%     \centering
%     \includegraphics{Fig/errLORviz_r50.pdf}
%     \caption{Low Order Reconstruction (LOR) error map for the jet flow (left) and the channel flow - front region (right). The LOR is computed using a rank $r=50$. The average velocity field is represented by black vectors. For clarity, only one vector is plotted for every six in the jet flow and every three in the channel flow. }
%     \label{fig:maperrLOR}
% \end{figure*}

\section{Conclusions}
In this study, we investigated the potential of using neuromorphic \acf{EBV} cameras for data-efficient low-order coordinate estimation in turbulent flows. More specifically, the \acf{ROM} performance of pulsed \acf{EBIV} is compared to conventional \ac{PIV}, using synchronized measurements for comparison, in the cases of a submerged water jet and a channel air-flow with a square rib.

Our findings highlight that \ac{EBIV} successfully captures flow statistics and dominant spectral content. Although the data exhibit higher noise compared to \ac{PIV}, leading to increased total energy and discrepancies in high-frequency spectral density, \ac{EBIV} reliably identifies the same dominant flow structures and dynamics as \ac{PIV}, including their associated energy. 
The distortion of low-order modes — critical for control — remains minimal, with noise primarily affecting higher-order modes by elevating their energy. More specifically, in the water jet case, slightly higher mode energy is associated with the first few modes. However, the cumulative mode energy aligns well with PIV due to comparable noise levels in the higher-order modes. In the channel air-flow case, the most relevant modes are well reconstructed, although a significantly higher cumulative energy is observed, attributed to the elevated noise in the higher-order modes.
Finally, \acf{LOR} demonstrate performance comparable to conventional \ac{PIV} measurements, even at relatively high-order modes. This confirms \ac{EBIV}'s potential to provide a reliable foundation for ROM of flow dynamics.
The higher sensitivity and lower data rate of \ac{EBV} cameras make them particularly attractive for real-time applications requiring low-order representations, such as predictive modeling and flow control. Nonetheless, challenges related to noise and event processing need to be addressed before effective control can be achieved.

It is important to emphasize that conventional PIV algorithms and experimental setups have been implemented, albeit applied to a novel technology. The authors believe that developing a dedicated event-based data-processing framework and fine-tuning the experimental setup—such as optimizing seeding density—can significantly enhance measurement quality.
Furthermore, the authors would like to highlight that multi-frame processing greatly improves the quality of the results, particularly for \ac{EBIV}. However, keeping in mind the goal of evaluating the real-time applicability of this technology, only simple dual-frame processing has been employed. %The true potential of EBIV, especially in high-speed velocimetry, lies in its ability to handle a continuous stream of data, which is well-suited for tracking and multi-frame processing, despite the potential loss of some information (e.g., less bright particles).

In conclusion, this study highlights the promising capabilities of \ac{EBV} cameras in the field of flow diagnostics and control. Despite challenges such as noise levels and event rate constraints, the significant reduction in data volume, enhanced sensor sensitivity, and innovative data representation position \ac{EBIV} as a strong candidate for real-time, imaging-based applications.

Future efforts should focus on developing more efficient data processing methods, integrating them directly into \ac{EBV} camera electronics, and improving robustness to latency. These advancements will be crucial for the deployment of \ac{EBIV} in real-time, closed-loop flow control systems.

\section*{CRediT author statement:} 
LF: Conceptualization, Data curation, Formal analysis, Investigation, Methodology, Software, Validation, Visualization, Writing - Original Draft, Writing - Review \& Editing; 
CW: Conceptualization,  Data curation, Investigation, Methodology, Resources, Software, Supervision, Validation, Writing - Review \& Editing
MR: Methodology, Supervision, Validation, Writing – review \& editing; 
SD: Conceptualization, Funding acquisition, Investigation, Methodology, Project administration, Resources, Supervision, Validation, Writing – review \& editing.

\section*{Declaration of competing interest}
The authors declare that they have no known competing financial interests or personal relationships that could have appeared to influence the work reported in this paper.

\section*{Data availability}
All datasets used in this work are openly available in Zenodo, accessible at \href{https://zenodo.org/uploads/14673928}{https://zenodo.org/uploads/14673928}. 

\section*{Acknowledgements}
This project has received funding from the European Research Council (ERC) under the European Union’s Horizon 2020 research and innovation programme (grant agreement No 949085, NEXTFLOW ERC StG). Views and opinions expressed are however those of the authors only and do not necessarily reflect those of the European Union or the European Research Council. Neither the European Union nor the granting authority can be held responsible for them.

The help of Michael Schroll of the DLR Institute of Technology was invaluable in setting up the air flow experiment.

\section*{Declaration of generative AI and AI-assisted technologies in the writing process}
During the preparation of this work, the authors used Chat-GPT and Grammarly to improve the readability and language of this manuscript. After using this tool/service, the authors reviewed and edited the content as needed and took full responsibility for the publication’s content.

%\todo{add DOIs in the refs where possible (Chris will take of it --> done)}
% \bibliographystyle{elsarticle-harv} 
% \bibliography{bibliography}

\end{document}